\documentclass[10pt]{article}
\usepackage{typearea}\typearea{14}
\usepackage{epsfig,amsmath,amsfonts,amssymb,upgreek,dsfont}
\newtheorem{theorem}{Theorem}[section]
\newtheorem{lemma}[theorem]{Lemma}
\newtheorem{corollary}[theorem]{Corollary}

\newtheorem{definition}[theorem]{Definition}

\makeatletter
\@addtoreset{equation}{section}
\makeatother

\makeatletter
\long\def\@makecaption#1#2{{\small
\advance\leftskip1cm
\advance\rightskip1cm
\vskip\abovecaptionskip
\sbox\@tempboxa{#1: #2}%
\ifdim \wd\@tempboxa >\hsize
 #1: #2\par
\else
\global \@minipagefalse
\hb@xt@\hsize{\hfil\box\@tempboxa\hfil}%
\fi
\vskip\belowcaptionskip}}
\makeatother
\def\eq#1\en{\begin{equation}#1\end{equation}}  
\def\eqa#1\ena{\begin{align}#1\end{align}}
\def\eqg#1\eng{\begin{gather}#1\end{gather}}
\newcommand{\lb}[1]{\label{e:#1}}
\newcommand{\rlb}[1]{\eqref{e:#1}} 
\newcommand{\nl}{\notag\\}

\newcommand{\qedm}{\rule{1.5mm}{3mm}}
\newcommand{\snorm}[1]{\Vert#1\Vert}

\newcommand{\sbkt}[1]{\langle#1\rangle}

\newcommand{\bra}[1]{\langle#1|}
\newcommand{\ket}[1]{|#1\rangle}

\newcommand{\sumtwo}[2]%
{\mathop{\sum_{#1}}_{#2}}
\newcommand{\sumthree}[3]%
{\mathop{\mathop{\sum_{#1}}_{#2}}_{#3}}
\newcommand{\sumfour}[4]%
{\mathop{\mathop{\mathop{\sum_{#1}}_{#2}}_{#3}}_{#4}} 
\newcommand{\prodtwo}[2]%
{\mathop{\prod_{#1}}_{#2}}
\newcommand{\mintwo}[2]%
{\mathop{\min_{#1}}_{#2}}
\newcommand{\maxtwo}[2]%
{\mathop{\max_{#1}}_{#2}}
\newcommand{\maxthree}[3]%
{\mathop{\mathop{\max_{#1}}_{#2}}_{#3}}
\newcommand{\limtwo}[2]%
{\mathop{\lim_{#1}}_{#2}}
\newcommand{\suptwo}[2]%
{\mathop{\sup_{#1}}_{#2}}
\newcommand{\supthree}[3]%
{\mathop{\mathop{\sup_{#1}}_{#2}}_{#3}}
\newcommand{\supfour}[4]%
{\mathop{\mathop{\mathop{\sup_{#1}}_{#2}}_{#3}}_{#4}} 
\newcommand{\inftwo}[2]%
{\mathop{\inf_{#1}}_{#2}}
\newcommand{\infthree}[3]%
{\mathop{\mathop{\inf_{#1}}_{#2}}_{#3}}
\newcommand{\inffour}[4]%
{\mathop{\mathop{\mathop{\inf_{#1}}_{#2}}_{#3}}_{#4}} 
\newcommand{\bsu}{\boldsymbol{u}}

\newcommand{\bsS}{\boldsymbol{S}}

\newcommand{\hc}{\hat{c}}
\newcommand{\hcd}{\hat{c}^\dagger}

\newcommand{\hh}{\hat{h}}

\newcommand{\hv}{\hat{v}}
\newcommand{\hn}{\hat{n}}

\newcommand{\hA}{\hat{A}}
\newcommand{\hB}{\hat{B}}

\newcommand{\hH}{\hat{H}}

\newcommand{\hN}{\hat{N}}

\newcommand{\hV}{\hat{V}}
\newcommand{\hW}{\hat{W}}
\newcommand{\hX}{\hat{X}}
\newcommand{\bbC}{\mathbb{C}}

\newcommand{\bbN}{\mathbb{N}}

\newcommand{\bbR}{\mathbb{R}}
\newcommand{\bbZ}{\mathbb{Z}}
\newcommand{\ep}{\varepsilon}

\newcommand{\up}{\uparrow}
\newcommand{\dn}{\downarrow}

\newcommand{\Di}{\mathit{\Delta}}
\newcommand{\DE}{{\Di E}}
\newcommand{\Tr}{\operatorname{Tr}}

\newcommand{\hbS}{\hat{\bsS}}
\newcommand{\hSx}{\hat{S}^{\rm x}}
\newcommand{\hSy}{\hat{S}^{\rm y}}
\newcommand{\hSz}{\hat{S}^{\rm z}}
\newcommand{\hU}{\hat{U}}
\newcommand{\vac}{\ket{\Phi_{\rm vac}}}

\newcommand{\GS}{\ket{\Phi_{\rm GS}}}
\newcommand{\GSb}{\bra{\Phi_{\rm GS}}}

\newcommand{\Hhop}{\hH_{\rm hop}}
\newcommand{\Hint}{\hH_{\rm int}}
\newcommand{\iop}{\hat{1}}
\newcommand{\Gphs}{\Gamma_{\rm phg}}
\newcommand{\Gph}{\Gamma_{\rm ph}}
\newcommand{\Gbi}{\Gamma_{\rm bi}}
\newcommand{\Gsi}{\Gamma_{\rm si}}
\newcommand{\Ind}{\mathrm{Ind}}
\newcommand{\LaL}{\Lambda_L}
\newcommand{\uds}{\{\up,\dn\}}
\newcommand{\Uone}{\mathrm{U}(1)}
\newcommand{\oA}{\mathfrak{A}}
\newcommand{\Aloc}{\mathfrak{A}_{\rm loc}}

\usepackage{color}
\usepackage[normalem]{ulem}
\definecolor{fluorescentpink}{rgb}{1.0, 0.08, 0.58}
\definecolor{forestgreen}{rgb}{0.13, 0.55, 0.13}

\begin{document}
\renewcommand{\thefootnote}{\fnsymbol{footnote}}

\noindent
{\Large\bf Rigorous Index Theory for One-Dimensional Interacting Topological Insulators}

\medskip\noindent
Hal Tasaki\footnote{%
Department of Physics, Gakushuin University, Mejiro, Toshima-ku, 
Tokyo 171-8588, Japan.
}
\renewcommand{\thefootnote}{\arabic{footnote}}
\setcounter{footnote}{0}

%%%%%%%%%%%%%%%%%
\begin{quotation}
\small\noindent
We present a rigorous but elementary index theory for a class of one-dimensional systems of interacting (and possibly disordered) fermions with $\Uone\rtimes\bbZ_2$ symmetry defined on the infinite chain.
The class includes the Su-Schrieffer-Heeger (SSH) model \cite{SSH1,SSH2,AOP} as a special case.
For any locally-unique gapped (fixed-charge) ground state of a model in the class, we define a $\bbZ_2$ index in terms of the sign of the expectation value of the local twist operator.
We prove that the index is topological in the sense that it is invariant under continuous modification of models in the class with a locally-unique (fixed-charge) gapped ground state.
This establishes that any path of models in the class that connects the two extreme cases of the SSH model must go through a phase transition.
Our rigorous $\bbZ_2$ classification is believed to be optimal for the class of models considered here.
We also show an interesting duality of the index, and prove that any topologically nontrivial model in the class has a gapless edge excitation above the ground state when defined on the half-infinite chain.
The results extend to other classes of models, including the extended Hubbard model.
Our strategy to focus on the expectation value of local unitary operators makes the theory intuitive and conceptually simple.
The paper also contains a careful discussion about the notion of unique gapped ground states of a particle system on the infinite chain.
(There are two lecture videos in which the main results of the paper are discussed \cite{video:short,video:long}.)
\end{quotation}

%%%%%%%%%%%%%%%%%%%%%%%%%%%%%%%%%%%%%
%%%%%%%%%%%%%%%%%%%%%%%%%%%%%%%%%%%%%

\tableofcontents

\section{Introduction}
\label{s:intro}
Topological insulators \cite{HasnKane} for non-interacting fermions are completely classified according to the  ``periodic table'' \cite{Ryu,Kitaev}, and are characterized by the indices that are written as an integral over the Brillouin zone when the model has translation invariance (see, e.g., \cite{HasnKane,Shankar}), or, more generally, by the indices for projections defined by methods of noncommutative geometry  (see, e.g., \cite{PSB,KatsuraKoma}).
Although similar classifications and characterizations of interacting topological insulators have been investigated intensively (see, e.g, \cite{Rachel,Hatsugai2006,GuoShen,FK,ShiozakiShapourianRyu,Manmana,WangXuWangWu2015,Stehouwer,Kapustin,Lu,Matsugatani,Ono,Kang,Wheeler,Nakamura}), mathematically rigorous results are still limited \cite{AvronSeiler,Bachmann1,Bachmann2,BSB,Matsui2020,BourneOgata,Ogata4}.
See \cite{BachmannNachtergaele2014A,Tasaki2018,Ogata1,Ogata2,OTT,Ogata3,Sopenko} for closely related rigorous index theorems for bosonic (or quantum spin) systems.

In the present paper, we develop an index theory for a class of one-dimensional interacting fermion systems that includes the Su-Schrieffer-Heeger (SSH) model \cite{SSH1,SSH2,AOP} as a special case.
As far as we understand our results do not follow from those in the literature \cite{AvronSeiler,Bachmann1,Bachmann2,BSB,Matsui2020,BourneOgata,Ogata4}.
The main object in our index theory is the ground state expectation value of the local twist operator.
The strategy to focus on {\em local}\/ operators in the {\em infinite}\/ system makes the theory intuitive and conceptually simple.
Although we here develop a fully rigorous index theory in the infinite chain, all the essential arguments can be intuitively understood within the language of standard quantum mechanics in finite systems.
See also the lecture videos in which the main results of the paper are discussed \cite{video:short,video:long}.

\begin{figure}
\centerline{\epsfig{file=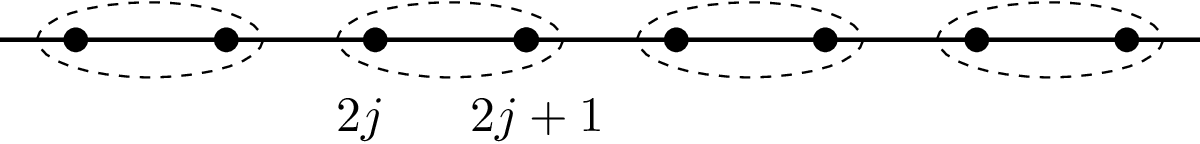,width=10truecm}}
\caption[dummy]{
A part of the infinite chain $\bbZ$.
We regard neighboring sites $2j$ and $2j+1$ (where $j\in\bbZ$) as forming a unit cell.
The assignment of unit cells is crucial throughout the present paper.
In the twist operator \rlb{U1}, two sites in a unit cell have the same twist angle.
In the extended Hubbard model \rlb{exHub}, two sites in a unit cell are coupled by the ferromagnetic interaction and behave as a single quantum spin with $S=1$.
}
\label{f:chain}
\end{figure}

\medskip

The SSH model \cite{SSH1,SSH2}, which is often discussed as a toy example of a topological insulator, is a non-interacting fermion model at half filling with the Hamiltonian
\eq
\hH^{\rm SSH}_s=\sum_j \bigl\{(1-s)(\hcd_{2j}\hc_{2j+1}+\text{h.c.})+s(\hcd_{2j-1}\hc_{2j}+\text{h.c.})\bigr\},
\lb{HSSH}
\en
where $s\in[0,1]$ is the model parameter.
We here regard sites $2j$ and $2j+1$ to form a unit cell.  See Figure~\ref{f:chain}.
The model has a unique gapped ground state unless $s=1/2$.
The two regions $[0,1/2)$ and $(1/2,1]$ of the parameter space separated by the gapless point $s=1/2$ are regarded as two distinct ``topological'' phases.
These phases are not characterized by standard order parameters but can be distinguished by the Zak phase \cite{Zak}, i.e., the Berry phase defined on the Brillouin zone
\eq
\nu=\frac{i}{\pi}\int_0^{2\pi}dk\,\sbkt{\bsu^-(k),\frac{d}{dk}\bsu^-(k)},
\en
which takes the values 0 and 1 when $s\in[0,1/2)$ and $s\in(1/2,1]$, respectively.
Here $\bsu^-(k)$ is the Bloch function for the lower band (see \cite{video:long} for the notation).
It is also known, in particular from the theory of polarization \cite{Resta94,Resta98,WatanabeOshikawa}, that the Zak phase is related to the ground state expectation value as
\eq
(-1)^\nu\simeq\bra{\Phi_{\rm GS}^{(L)}}\hU_L\ket{\Phi_{\rm GS}^{(L)}}.
\en
Here 
\eq
\hU_L=\exp\Bigl[i\sum_{j=1}^{L/2}\frac{4\pi j}{L}(\hn_{2j}+\hn_{2j+1}-1)\Bigr],
\en
is the twist operator of Bloch \cite{Bohm,Watanabe} and Lieb, Schultz, and Mattis \cite{LiebSchultzMattis1961,YOA}, and $\ket{\Phi_{\rm GS}^{(L)}}$ denotes the ground state of a finite periodic chain with $L$ sites.
See \cite{video:long} for a review.

It is worth noting that, unlike the Chern number, the Zak phase is not always quantized.
Here it is quantized to take only two values, 0 or 1, because the model possesses the $\bbZ_2$ symmetry discussed in section~\ref{s:Z2} in addition to the U(1) symmetry intrinsic to quantum particle systems.
In this sense, the topological phases of the SSH model should be called symmetry-protected topological (SPT) phases.

We shall study interacting versions of the SSH model with a possible disorder, obtained by adding arbitrary real hopping between odd and even sites and arbitrary interaction of the form $\sum_{j,k}v_{j,k}(\hn_j-\frac{1}{2})(\hn_k-\frac{1}{2})$ to the original Hamiltonian \rlb{HSSH}.
It has been argued that topological phases of the interacting SSH (and related) models are characterized by the Berry phase (defined for a family of ground states with twisted boundary conditions) \cite{Hatsugai2006,GuoShen,ShiozakiShapourianRyu}, which should be related to the expectation value of the twist operator \cite{WatanabeOshikawa}. (See endnote~30 of \cite{Tasaki2018} for details.)

Our index theory can be regarded as a rigorous version of the above heuristic approaches.
Since phase transitions must be ultimately discussed in infinite systems, we define a $\bbZ_2$-valued index for a locally-unique gapped (fixed-charge) ground state on the infinite chain by using the local twist operator introduced by Affleck and Lieb \cite{AffleckLieb1986,Tasaki2017A} (Theorem~\ref{t:index}, Definition~\ref{d:index}), and prove that it is ``topological'', i.e., invariant under a continuous modification of locally-unique gapped ground states (Corollary~\ref{c:index}).
With the index in hand, we prove that a topologically nontrivial model defined on the half-infinite chain inevitably has a gapless excitation near the edge (Theorems~\ref{t:edge1} and \ref{t:edge2}).
We also state an interesting duality of the index and prove that any locally-unique gapped ground state in this class of model is indeed topologically nontrivial if one chooses the unit cells appropriately (Theorem~\ref{t:duality}).

The present theory is an improvement and extension of our earlier work \cite{Tasaki2018} (based on the observation in \cite{NakamuraTodo2002}) for quantum spin chains. (See, e.g., \cite{TasakiBook} for background.)
Our theory is much more satisfactory in the present setting since U(1) symmetry, which is essential for the proof, is inherent in quantum particle systems.
We indeed get a classification that is believed to be optimal.

\medskip
The present paper is organized as follows.
In section~\ref{s:SSH}, we define the SSH-type models of interacting fermions on the infinite chain.
We carefully discuss the definitions of ground states in infinite systems.
We present our rigorous index theory for the SSH-type models in section~\ref{s:Indextheory}.
All the theorems are then proved in section~\ref{s:proof}.
In section~\ref{s:other}, we briefly discuss extensions to other important classes of models, including the extended Hubbard model.

%%%%%%
\section{The SSH-type models and their ground states}\label{s:SSH}
We introduce the most basic class of models we study, namely, the SSH-type models, and discuss their symmetry.
We also present a general discussion about the definitions of ground states on the infinite chain.

\subsection{General model with U(1) symmetry}
\label{s:U1}
We consider a system of spinless fermions on the infinite chain $\bbZ$, and denote by $\hcd_j$ and $\hc_j$ the standard creation and annihilation operators at site $j\in\bbZ$.
They satisfy the canonical anticommutation relations
\eq
\{\hc_j,\hc_k\}=0,\quad\{\hc_j,\hcd_k\}=\delta_{j,k},
\lb{CAC}
\en
for any $j,k\in\bbZ$.
(Here $\{\hA,\hB\}$ stands for $\hA\hB+\hB\hA$.)
The number operator is defined by $\hn_j=\hcd_j\hc_j$.
Let $\Aloc$ be the set of all local operators, i.e., polynomials of $\hc_j$ and $\hcd_j$ with $j\in\bbZ$.
(Note that a local operator acts nontrivially on an arbitrary finite number of sites.)

We take the standard Hamiltonian 
\eq
\hH=\Hhop+\Hint,
\lb{H0}
\en
where $\Hhop$ and $\Hint$ are the hopping Hamiltonian and the interaction Hamiltonian, respectively.
We formally write them as infinite sums
\eq
\Hhop=\sum_{j,k\in\bbZ}t_{j,k}\,\hcd_j\hc_k,\quad
\Hint=\sum_{j\in\bbZ}\hv_j,
\lb{H1}
\en
where $(t_{j,k})^*=t_{k,j}\in\bbC$ and $\hv_j^\dagger=\hv_j$.
We assume that the Hamiltonian is short-ranged in the sense that there is a constant $r_0$ such that $t_{j,k}=0$ whenever $|j-k|>r_0$, and $\hv_j$ is a polynomial of $\hn_k$ with $|k-j|\le r_0$.
We also assume that there are constants $t_0$, $v_0$ such that 
\eq
\frac{1}{2}\sum_{k\in\bbZ\backslash\{j\}}|t_{j,k}|(|k-j|+1)^2\le t_0,\quad\snorm{\hv_j}\le v_0
\lb{tshort}
\en
hold for any $j$.\footnote{%
Throughout the present paper, $\snorm{\hA}$ denotes the operator norm of $\hA$.
}
Note that neither $\Hhop$ nor $\Hint$ is an element of $\Aloc$.

The most crucial symmetry is the quantum mechanical U(1) symmetry which is already built into our models.
From the above definitions, we see that our Hamiltonian conserves the total particle number defined formally as
\eq
\hN=\sum_{j\in\bbZ}\hn_j.
\lb{hN}
\en
In other words we have the commutation relation $[t_{j,k}\,\hcd_j\hc_k,\hN]=0$ and $[\hv_j,\hN]=0$ for any $j,k\in\bbZ$.
To be rigorous, we define the commutator of a local operator $\hA\in\Aloc$ and the total number operator $\hN$, which is only formally defined, by
\eq
[\hA,\hN]:=\Bigl[\hA,\sum_{j\in I}\hn_j\Bigr],
\lb{AN}
\en
where the interval $I\subset\bbZ$ is chosen so as to include the support of $\hA$.
Clearly, the right-hand side of \rlb{AN} is independent of the choice of $I$ (provided that $I$ includes the support).

The above-mentioned conservation law implies for any $\theta\in\bbR$ that
\eqg
\exp\Bigl[-i\theta\sum_{m\in I}\hn_m\Bigr]\,t_{j,k}\,\hcd_j\hc_k\,\exp\Bigl[i\theta\sum_{m\in I}\hn_m\Bigr]=t_{j,k}\,\hcd_j\hc_k,\lb{U11}\\
\exp\Bigl[-i\theta\sum_{m\in I}\hn_m\Bigr]\,\hv_j\,\exp\Bigl[i\theta\sum_{m\in I}\hn_m\Bigr]=\hv_j,
\lb{U12}
\eng
where the interval $I$ is chosen to include the support of the relevant operators.
These represent the U(1) invariance of the model.

%%%
\subsection{States and ground states}
We shall make clear what we mean by a unique gapped ground state of a quantum system of particles defined on the infinite chain $\bbZ$.
An elementary introduction and further references for the case of quantum spin systems can be found in \cite{TasakiBook,TasakiLSM}.\footnote{%
The treatment of quantum spin systems is easier since we do not consider fixed-charge states.}
For our index theorem, a class of states that we call ``locally-unique gapped fixed-charge ground states'' is most relevant.
As we shall prove in section~\ref{s:IVL}, this notion is directly related to that of a unique gapped ground state on a finite chain, which is standard in the physics literature.

We start with the definition of states in general.
In infinite quantum systems, it is convenient to work with a linear map $\rho$ that gives the expectation value $\rho(\hA)$ of an operator $\hA\in\Aloc$.

\begin{definition}[state]
A state $\rho$ is a linear map from\footnote{%
To be rigorous, a state is a linear map from the C$^*$-algebra $\oA=\overline{\Aloc}$, the norm completion of $\Aloc$, to $\bbC$.
} $\Aloc$ to $\bbC$ such that $\rho(\iop)=1$ and $\rho(\hA^\dagger\hA)\ge0$ for any $\hA\in\Aloc$.
\end{definition}
These conditions imply $\rho(\hA^\dagger)=\rho(\hA)^*$ and $|\rho(\hA)|\le\snorm{\hA}$.
In a finite quantum system, a state $\rho$ is given by $\rho(\hA)=\bra{\Phi}\hA\ket{\Phi}$ with a normalized pure state $\ket{\Phi}$ or, more generally, by $\rho(\hA)=\Tr[\hA\hat{\rho}]$ with a density matrix $\hat{\rho}$.

We shall define some types of ground states.
Fixed-charge ground states are most relevant to us and are also most natural for a system that conserves the particle number.
As suggested by the terminology, it represents a ground state in a fixed-charge (i.e., fixed particle number) sector. 
The definition essentially says that one cannot lower the energy of the state $\omega$ by a particle number preserving perturbation $\hV$. 

\begin{definition}[fixed-charge ground state]\label{d:FCGS}
A state $\omega$ is said to be a fixed-charge ground state of $\hH$ if it holds that $\omega(\hV^\dagger[\hH,\hV])\ge0$ for any $\hV\in\Aloc$ such that $[\hV,\hN]=0$.
\end{definition}
Recall that the Hamiltonian $\hH$ is only defined formally by infinite sums \rlb{H1}.
But one can unambiguously define the commutator $[\hH,\hV]$ as in \rlb{AN} since $\hV$ is a local operator.

It is useful to note that, in a finite system, this definition reduces to the standard definition of ground states in the Hilbert space with a fixed particle number.
Let $\ket{\Phi}$ be a normalized pure state of a finite system and write $\omega(\cdot)=\bra{\Phi}\cdot\ket{\Phi}$.
Then the condition  $\omega(\hV^\dagger[\hH,\hV])\ge0$ reads $\bra{\Phi}\hV^\dagger\hH\hV\ket{\Phi}\ge\bra{\Phi}\hV^\dagger\hV\hH\ket{\Phi}$.
If we take $\ket{\Phi}$ as a ground state $\GS$ with energy $E_{\rm GS}$, this becomes $\bra{\Psi}\hH\ket{\Psi}\ge E_{\rm GS}$, 
where $\ket{\Psi}=\hV\GS/\snorm{\hV\GS}$ is a normalized variational state.
We thus get the standard variational characterization of a ground state.
When $\ket{\Phi}$ is not a ground state, one immediately sees that the condition is violated by taking $\hV=\GS\bra{\Phi}$.

Since we have considered only particle number preserving perturbations $\hV$, we expect that there is at least one fixed-charge ground state in each fixed-charge (or fixed particle number) sector (although we here do not try to make the notion of fixed-charge sectors rigorous).
It is also common to employ the Fock space formalism and define a ground state within the space of varying particle numbers.
In this case, we introduce a new parameter $\mu\in\bbR$, the (zero-temperature) chemical potential, and define 
\eq
\hH^{(\mu)}=\hH-\mu\hN,
\lb{Hmu}
\en
where $\hH$ and $\hN$ are defined as \rlb{H0} and \rlb{hN}, respectively.
We then define a ground state as follows.
\begin{definition}[ground state]\label{d:GS}
A state $\omega$ is said to be a ground state of $\hH^{(\mu)}$ if it holds that $\omega(\hV^\dagger[\hH^{(\mu)},\hV])\ge0$ for any $\hV\in\Aloc$.
\end{definition}
Note that the perturbation $\hV$ is completely arbitrary except for being local.
In this definition, the particle number (or, more precisely, the particle density at each site) is determined by the chemical potential $\mu$.
In fact, we only treat the case with $\mu=0$ for the SSH-type model, which has extra $\bbZ_2$ symmetry as discussed in section~\ref{s:Z2}.

If $\omega$ is a ground state of $\hH^{(\mu)}$ for some $\mu$, it is clearly a fixed-charge ground state of $\hH$.
The converse is not true in general.\footnote{%
We expect that a pure fixed-charge ground state $\omega$ of $\hH$ is a ground state of $\hH^{(\mu)}$ with suitable $\mu$, except at phase transition points.}

As in the simple example of the SSH model discussed in section~\ref{s:intro}, we are interested in a ground state that is unique and accompanied by a nonzero energy gap.
In an infinite system, a corresponding useful notion is known as a locally-unique gapped (fixed-charge) ground state.\footnote{%
This terminology was introduced recently in \cite{TasakiLSM}, but the notion itself was known before.
See, e.g., \cite{Nachtergaele2022}.}
We say that a  (fixed-charge) ground state $\omega$ is locally-unique and gapped if any local excitation (that preserves the particle number) costs excitation energy not less than a finite constant $\DE$.
Here $\DE$ is the energy gap above the ground state.
To be precise, this is defined as the following for the fixed-charge setting and the Fock space setting, respectively.
\begin{definition}[locally-unique gapped fixed-charge ground state]\label{d:LUGFCGS}
A fixed-charge ground state $\omega$ of $\hH$ is said to be locally-unique and gapped if there is a constant $\varepsilon>0$ such that $\omega(\hV^\dagger[\hH,\hV])\ge\varepsilon\,\omega(\hV^\dagger\hV)$ for any $\hV\in\Aloc$ that satisfies $[\hV,\hN]=0$ and $\omega(\hV)=0$.
The energy gap $\DE$ of $\omega$ is the largest $\varepsilon$ with the above property.
\end{definition}
\begin{definition}[locally-unique gapped ground state]\label{d:LUGGS}
A ground state $\omega$ of $\hH^{(\mu)}$ is said to be locally-unique and gapped if there is a constant $\varepsilon>0$ such that $\omega(\hV^\dagger[\hH^{(\mu)},\hV])\ge\varepsilon\,\omega(\hV^\dagger\hV)$ for any $\hV\in\Aloc$ that satisfies $\omega(\hV)=0$.
The energy gap $\DE$ of $\omega$ is the largest $\varepsilon$ with the above property.
\end{definition}
It is again apparent that if $\omega$ is a locally-unique gapped ground state of $\hH^{(\mu)}$ for some $\mu$, then it is also a locally-unique gapped fixed-charge ground state of $\hH$.\footnote{%
The converse is much more delicate in this case since a locally-unique gapped ground state should have a gap for all possible excitations, while a locally-unique gapped fixed-charge ground state is only required to have a gap for particle number preserving excitations.
}

To see the relation with the standard definition for a finite system, observe (in the fixed-charge setting) that the conditions $\omega(\hV^\dagger[\hH,\hV])\ge\varepsilon\,\omega(\hV^\dagger\hV)$ and $\omega(\hV)=0$ read  $\bra{\Psi}\hH\ket{\Psi}\ge E_{\rm GS}+\varepsilon$ and $\GSb\Psi\rangle=0$, respectively, for the same $\GS$ and $\ket{\Psi}$ as above.
These are precisely the variational characterization of a unique gapped ground state.

A crucial feature of the notion of locally-unique gapped (fixed-charge) ground states is that the infinite volume limit of a sequence of unique gapped ground states of finite systems automatically becomes a locally-unique gapped (fixed-charge) ground state.
This is true for both the fixed-charge setting and the Fock space setting.
See section~\ref{s:IVL} for details.
This implies, in particular, that the infinite volume limits of the ground states of the SSH model \rlb{HSSH} with $s\ne1/2$ are locally-unique gapped fixed-charge ground states. 

We finally discuss the most standard (and abstract) definition of the unique gapped ground state for an infinite system.
As far as we know, a general definition is possible only for the Fock space setting.
\begin{definition}[unique gapped ground state]
\label{d:UGGS}
A state $\omega$ is said to be a unique gapped ground state of $\hH^{(\mu)}$ if it is the only ground state of $\hH^{(\mu)}$ (in the sense of Definition~\ref{d:GS}) and is a locally-unique gapped ground state (in the sense of Definition~\ref{d:LUGGS}).
\end{definition}
If $\omega$ is a unique gapped ground state of $\hH^{(\mu)}$, it is automatically a locally-unique gapped ground state of $\hH^{(\mu)}$.
But the converse is not true in general.  See \cite{TasakiLSM}.

In section~\ref{s:Indextheory}, we shall state our index theorems for a locally-unique gapped fixed-charge ground state of $\hH$ (Definition~\ref{d:LUGFCGS}).
Since any unique gapped ground state (Definition~\ref{d:UGGS}) or locally-unique gapped ground state (Definition~\ref{d:LUGGS}) of $\hH^{(\mu)}$ is also a locally-unique gapped fixed-charge ground state of $\hH$, our theorems indeed apply to a unique gapped ground state according to any of the three definitions.

\subsection{The infinite volume limit of unique gapped ground states}
\label{s:IVL}
Let us provide the details about the relationship between unique gapped ground states in finite chains and locally-unique gapped (fixed-charge) ground states on the infinite chain.

Let $L$ be even and consider a system of fermions on a finite chain $\LaL=\{-L/2,\ldots,L/2\}\subset\bbZ$ with the Hamiltonian $\hH_L$ written as
\eq
\hH_L=\sum_{j,k\in\LaL}t_{j,k}\,\hcd_j\hc_k+\sumtwo{j\in\LaL}{(\operatorname{supp}\hv_j\subset\LaL)}\hv_j+\Di\hH_L,
\en
where the hopping $t_{j,k}$ and the interaction $\hv_j$ are the same as in the infinite volume Hamiltonian $\hH$ given by \rlb{H0} and \rlb{H1}.
(The second summation is restricted to the range where the support of $\hv_j$ is contained in $\LaL$.)
Here $\Di\hH_L$ is a suitable boundary Hamiltonian that acts on sites within a fixed distance (independent of $L$) from the two boundaries.
This, in particular, allows us to treat the periodic boundary condition.
In the SSH-type models, it is convenient to further assume that $\Di\hH_L$ has $\Uone\rtimes\bbZ_2$ symmetry.

In the fixed-charge setting, we fix particle number $N_L$ for each $L$, and work within the Hilbert space with $N_L$ particles.
We then assume for each $L$ that $\hH_L$ has a unique gapped ground state $\ket{\Phi_{\rm GS}^{(L)}}$ (in the sense of standard quantum mechanics) with energy gap not less than a constant $\DE>0$ independent of $L$.

In the Fock space setting, we fix the chemical potential $\mu$ and assume that the Hamiltonian 
\eq
\hH_L^{(\mu)}=\hH_L-\mu\sum_{j=1}^L\hn_j,
\en
has a unique gapped ground state $\ket{\Phi_{\rm GS}^{(L)}}$ (in the sense of standard quantum mechanics) with energy gap not less than a constant $\DE>0$ independent of $L$.
Here we consider the whole Hilbert space with any number of particles.

In both cases, we define a state $\omega$ on the infinite chain by
\eq
\omega(\hA)=\lim_{L\up\infty}\bra{\Phi_{\rm GS}^{(L)}}\hA\ket{\Phi_{\rm GS}^{(L)}},
\lb{lim1}
\en
for any $\hA\in\Aloc$.
Note that the expectation value $\bra{\Phi_{\rm GS}^{(L)}}\hA\ket{\Phi_{\rm GS}^{(L)}}$ is well-defined for sufficiently large $L$ since $\hA$ is local.
Of course the limit \rlb{lim1} may not exist.
It is known, however, that one can always take a subsequence, i.e., a strictly increasing function $L(n)\in\bbN$ of $n\in\bbN$, such that the limit
\eq
\omega(\hA)=\lim_{n\up\infty}\bra{\Phi_{\rm GS}^{(L(n))}}\hA\ket{\Phi_{\rm GS}^{(L(n))}},
\lb{lim2}
\en
exists for any $\hA\in\Aloc$.\footnote{%
This is a formal result that follows from an abstract theorem (the Banach-Alaoglu theorem) in functional analysis.
See, e.g., Theorem~A.24 (p.488) of \cite{TasakiBook} for more details.
}
We define the state $\omega$ on the infinite chain by the limit \rlb{lim2}.

Then we have the following desired result for the state $\omega$.
\begin{theorem}\label{t:finitetoinfinite}
The state $\omega$ is a locally-unique gapped fixed-charge ground state of $\hH$ in the fixed-charge setting, and is a locally-unique gapped ground state of $\hH^{(\mu)}$ in the Fock space setting.
\end{theorem}

\noindent
{\em Proof:}
We write $\omega_L(\cdot)=\bra{\Phi_{\rm GS}^{(L)}}\cdot\ket{\Phi_{\rm GS}^{(L)}}$.
Let $\hV\in\Aloc$ be an arbitrary local operator.
For the fixed-charge setting, we further assume that $[\hV,\hN]=0$.
Take sufficiently large $L$ such that the support of $\hV$ is contained in $\LaL$ and does not overlap with the support of $\Di\hH_L$.
We then have $[\hH_L,\hV]=[\hH,\hV]$.

On the other hand the standard variational principle implies $\omega_L(\hV^\dagger\,[\hH_L,\hV])\ge0$, which means  $\omega_L(\hV^\dagger\,[\hH,\hV])\ge0$.
By letting $L\up\infty$ (or, more precisely, $n\up\infty$ in $L(n)$) we see that $\omega$ is a (fixed-charge) ground state in the sense of Definitions~\ref{d:FCGS} or \ref{d:GS}.

To see that $\omega$ is locally-unique and gapped, we further assume that $\omega(\hV)=0$.
Let $\hV_L=\hV-\omega_L(\hV)$.
Note that this means $\hV_L\ket{\Phi_{\rm GS}^{(L)}}$ is orthogonal to $\ket{\Phi_{\rm GS}^{(L)}}$.
Since $\ket{\Phi_{\rm GS}^{(L)}}$ is unique and gapped, we find from the variational principle that
\eq
\omega_L(\hV_L^\dagger\,[\hH_L,\hV_L])\ge\DE\,\omega_L(\hV_L^\dagger\,\hV_L).
\en
This means, for sufficiently large $L$, that
\eq
\omega_L(\hV_L^\dagger\,[\hH,\hV])\ge\DE\,\omega_L(\hV_L^\dagger\,\hV_L).
\en
Since $\hV_L\to\hV$ as $L\up\infty$, we get the desired condition
\eq
\omega(\hV^\dagger\,[\hH,\hV])\ge\DE\,\omega(\hV^\dagger\,\hV),
\en
for a locally-unique gapped (fixed-charge) ground state as in Definitions~\ref{d:LUGFCGS} or \ref{d:LUGGS}.~\qedm

\subsection{$\bbZ_2$ symmetry}
\label{s:Z2}
The SSH-type models are characterized by a discrete $\bbZ_2$ symmetry in addition to the U(1) symmetry discussed in section~\ref{s:U1}.
The models thus have $\Uone\rtimes\bbZ_2$ symmetry.

To specify the symmetry, we introduce the notion of linear $*$-automorphism.
A one-to-one linear map $\Gamma$ from $\Aloc$ to itself is said to be a linear $*$-automorphism if 
\eq
\Gamma(\hA\hB)=\Gamma(\hA)\Gamma(\hB),\quad\Gamma(\hA^\dagger)=\Gamma(\hA)^\dagger
\en
for any $\hA,\hB\in\Aloc$.
The following definition of invariance should be obvious.
\begin{definition}
Let $\Gamma$ be a linear $*$-automorphism on $\Aloc$ and $\rho$ be a state.
We say $\rho$ is invariant under $\Gamma$, or $\Gamma$-invariant, if $\rho(\Gamma(\hA))=\rho(\hA)$ for any $\hA\in\Aloc$.
\end{definition}

In the SSH-type models, we assume the symmetry described by the linear $*$-automorphism $\Gphs$ 
defined by 
\eq\Gphs(\hc_j)=(-1)^j\,\hcd_j,
\en
for any $j\in\bbZ$.
It represents the combination of the particle-hole transformation $\hc_j\to\hcd_j$ on all sites and the gauge transformation $\hcd_j\to-\hcd_j$ on odd sites.
Note, in particular, that
\eq
\Gphs(\hn_j)=\Gphs(\hcd_j)\Gphs(\hc_j)=(-1)^{2j}\hc_j\hcd_j=1-\hn_j.
\en
We note in passing that in a finite chain, $\Gphs$ is realized as $\Gphs(A)=\hU_{\rm phg}^\dagger\hA\hU_{\rm phg}$ with a unitary operator 
\eq
\hU_{\rm phg}=\prod_{j:\text{odd}}(\hc_j-\hcd_j)\prod_{j:\text{even}}(\hc_j+\hcd_j).
\en

We require that the Hamiltonian $\hH$ defined in \rlb{H0} and \rlb{H1} is invariant under $\Gphs$.
To be precise, we require that the Hamiltonian is reorganized as $\hH=\sum_{j\in\bbZ}\hh_j$ with suitable $\hh_j\in\Aloc$ such that $\Gphs(\hh_j)=\hh_j$.
This is realized by requiring that $t_{j,k}$ is real if $j-k$ is odd, $t_{j,k}$ is pure imaginary if $j-k$ is even, and $t_{j,j}=0$, and that $\hv_j$ is an even polynomial of $(n_k-\frac{1}{2})$.
Note that, when $\hH$ is $\Gphs$-invariant, the Hamiltonian $\hH^{(\mu)}$ of \rlb{Hmu} is $\Gphs$-invariant only when $\mu=0$.

\section{Index theory}
\label{s:Indextheory}
We are now ready to discuss our main results.
The proofs, although elementary and not too long, will be presented separately in section~\ref{s:proof}.

We shall study a $\Gphs$-invariant locally-unique gapped fixed-charge ground state of an SSH-type model defined in the previous section.
The SSH model \cite{SSH1,SSH2,AOP} at half-filling discussed in section~\ref{s:intro} is a special case.
These models belong to class D in the classification of topological insulators\footnote{
There are two different definitions of class D for interacting fermions.  See, e.g., \cite{Stehouwer}.
Mathematically rigorous $\bbZ_2$ index theories in \cite{BSB,Matsui2020} apply to class D according to the other definition.
It is interesting that our index theory has some similarities with those in \cite{BSB,Matsui2020}.
}, where the topological phases are expected to be indexed by $\bbZ_2$ \cite{ShiozakiShapourianRyu,Stehouwer}.
This expectation suggests that our rigorous index theory provides a full classification of the topological phases in these models.

\subsection{Main index theorem}
Consider a short-ranged $\Gphs$-invariant Hamiltonian $\hH$ defined as in \rlb{H0} and \rlb{H1}.

We take the fixed-charge setting and assume that $\hH$ has a locally-unique gapped fixed-charge ground state $\omega$.
We also make a crucial assumption that $\omega$ is invariant under $\Gphs$, i.e., it satisfies\footnote{%
Note that a locally-unique gapped (fixed-charge) ground state of a $\Gphs$-invariant Hamiltonian is not necessarily $\Gphs$-invariant.
}
\eq
\omega(\Gphs(\hA))=\omega(\hA),
\lb{oInv}
\en 
for any $\hA\in\Aloc$.
The invariance implies $\omega(\hn_j)=\omega(\Gphs(\hn_j))=1-\omega(\hn_j)$ and hence $\omega(\hn_j)=1/2$.
We see that the system is at half-filling.

If the Fock-space setting is preferable, we assume that the Hamiltonian\footnote{
\label{fn:H0}
To be consistent in notation, $\hH$ should better be $\hH^{(0)}$, i.e., \rlb{Hmu} with $\mu=0$.
} $\hH$ has a locally-unique gapped ground state $\omega$, and also that $\omega$ is $\Gphs$-invariant.
As we noted before, such $\omega$ is also a  locally-unique gapped fixed-charge ground state of the Hamiltonian $\hH$.

\begin{figure}
\centerline{\epsfig{file=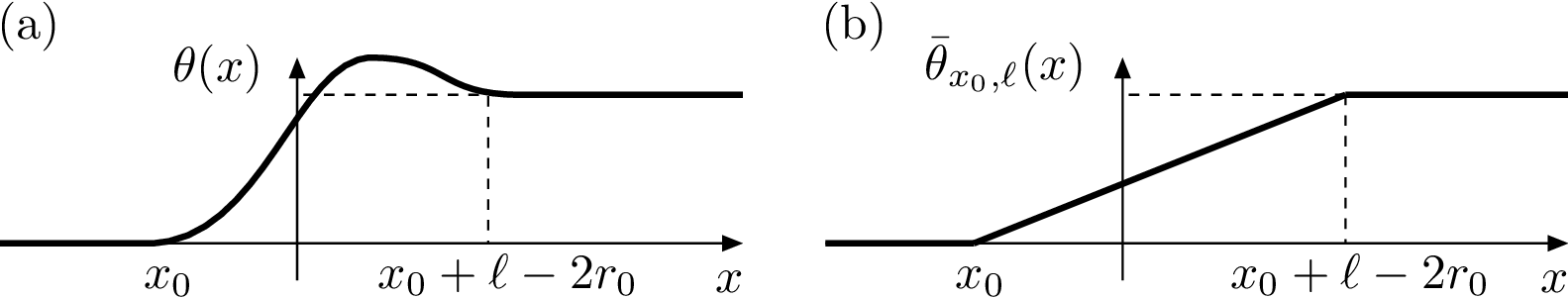,width=14truecm}}
\caption[dummy]{
(a) A general $\theta$-function. (b)~The piecewise linear $\theta$-function \rlb{bartheta}.
}
\label{f:theta}
\end{figure}

To define the index, we represent the unit circle $S^1$ as the interval $[0,2\pi]$ where 0 and $2\pi$ are identified.
We take a continuous and piecewise smooth function $\theta:\bbR\to S^1$ such that
\eq
\theta(x)=
\begin{cases}
0,&x\in(-\infty,x_0];\\
2\pi,&x\in[x_0+\ell-2r_0,\infty),
\end{cases}
\lb{thetacond}
\en
for some $x_0$ and $\ell$.
We assume that there is a constant $\gamma>0$ such that $|\theta'(x)|\le\gamma$ whenever the derivative $\theta'(x)$ exists.
We finally require that $\theta(x)$ wraps around $S^1$ once in the positive direction as $x$ varies from $x_0$ to $x_0+\ell-2r_0$.
See Figure~\ref{f:theta}~(a).
The simplest choice is the piecewise linear function
\eq
\bar{\theta}_{x_0,\ell}(x)=\begin{cases}
0,&x\in(-\infty,x_0];\\
\gamma\,(x-x_0),&x\in[x_0,x_0+\ell-2r_0];\\
2\pi,&x\in[x_0+\ell-2r_0,\infty),
\end{cases}
\lb{bartheta}
\en
with $\gamma=2\pi/(\ell-2r_0)$.
See  in Figure~\ref{f:theta}~(b).
In fact, we can develop our index theory by only using the function \rlb{bartheta}.
We shall however use general $\theta$-function to stress the robustness of our definition of the index.

Following \cite{Bohm,Watanabe,LiebSchultzMattis1961,YOA,AffleckLieb1986,Tasaki2017A}, we define the local twist operator or the flux-insertion operator $\hU_\theta$ by
\eq
\hU_\theta=\exp\Bigl[i\sum_{j\in\bbZ}\theta(2j)\,(\hn_{2j}+\hn_{2j+1}-1)\Bigr].
\lb{U1}
\en
Because of the crucial relation $\exp[i2\pi(\hn_{2j}+\hn_{2j+1}-1)]=1$, the summation can be restricted to $j$ such that $2j\in[x_0,x_0+\ell-2r_0]$.
This means that $\hU_\theta$ is a local operator.
Note also that the definition \rlb{U1} reflects our convention to regard sites $2j$ and $2j+1$ as forming a unit cell.
See Figure~\ref{f:chain}.

Noting that $\Gphs(\hn_{2j}+\hn_{2j+1}-1)=-(\hn_{2j}+\hn_{2j+1}-1)$, we see that
$\Gphs(\hU_\theta)=\hU_\theta^\dagger$,
which, with the invariance \rlb{oInv} of $\omega$, implies that $\omega(\hU_\theta)=\omega(\hU_\theta)^*$ and  hence 
\eq
\omega(\hU_\theta)\in\bbR.
\en
The reality of the expectation value $\omega(\hU_\theta)$ plays an important role.
The following theorem is the basis of our index theory.
\begin{theorem}
\label{t:index}
Assume that $\hH$ has a $\Gphs$-invariant locally-unique gapped fixed-charge ground state $\omega$ with energy gap $\DE$.
Then, for any $\theta$-function characterized by $\gamma$ and $\ell$ such that $\gamma^2\ell<\DE/t_0$, the expectation value $\omega(\hU_\theta)$ is nonzero, and its sign is independent of the choice of $\theta$ (satisfying the condition $\gamma^2\ell<\DE/t_0$).
\end{theorem}

Note that, for the piecewise linear function \rlb{bartheta}, we have
\eq
\gamma^2\ell=4\pi^2\frac{\ell}{(\ell-2r_0)^2},
\en
which can be made as small as one wishes by letting $\ell$ large.
This means that the condition $\gamma^2\ell<\DE/t_0$ can always be satisfied for some $\theta$-function.\footnote{%
Note that this is true only for the infinite chain.
In this sense, our index theory relies essentially on the fact that the chain is infinite.
}

The theorem and the above observation enable us to define a $\bbZ_2$-valued index $\Ind_\omega\in\{1,-1\}$ that characterizes the ground state $\omega$ as follows.
\begin{definition}\label{d:index}
Let $\omega$ be a $\Gphs$-invariant locally-unique gapped fixed-charge ground state.
We define
\eq
\Ind_\omega=\begin{cases}
1&\text{\em if $\omega(\hU_\theta)>0$};\\
-1&\text{\em if $\omega(\hU_\theta)<0$},
\end{cases}
\en
for any $\theta$-function satisfying the condition in Theorem~\ref{t:index}.
We say a ground state $\omega$ with $\Ind_\omega=1$ is topologically trivial and that with $\Ind_\omega=-1$ is topologically nontrivial.
\end{definition}

We can also show that the index is ``topological'' in the sense that it is invariant under continuous modification of $\Gphs$-invariant locally-unique gapped fixed-charge ground states.
Consider a family of short-ranged  $\Gphs$-invariant Hamiltonians $\hH_s$ of the form \rlb{H0}, \rlb{H1} with $s\in[0,1]$.
We assume that, for each $s\in[0,1]$, the Hamiltonian $\hH_s$ has a  $\Gphs$-invariant locally-unique gapped fixed-charge ground state $\omega_s$ whose energy gap is lower bounded by a constant $\DE_0>0$.
We further assume that the ground states depend continuously on $s$, i.e., $\omega_s(\hA)$ is continuous in $s$ for any $\hA\in\Aloc$.
Then we have the following.

\begin{corollary}\label{c:index}
In the above setting, the index $\Ind_{\omega_s}$ is independent of $s$.
\end{corollary}

\noindent
{\em Proof:}
Fix a function $\theta$ that satisfies $\gamma^2\ell<\DE_0/t_0$.
Theorem~1 shows that $\omega_s(\hU_\theta)\ne0$ for any $s$.
Then it is apparent from continuity that $\omega_s(\hU_\theta)$ cannot change its sign.~\qedm

\medskip
We thus conclude that, whenever $\Ind_{\omega_0}\ne \Ind_{\omega_1}$, there must be a certain phase transition (where one has a gapless ground state, non-locally-unique ground states, or discontinuity) at an intermediate $s$ in the one-parameter family of $\Gphs$-invariant Hamiltonians.
.

\begin{figure}
\centerline{\epsfig{file=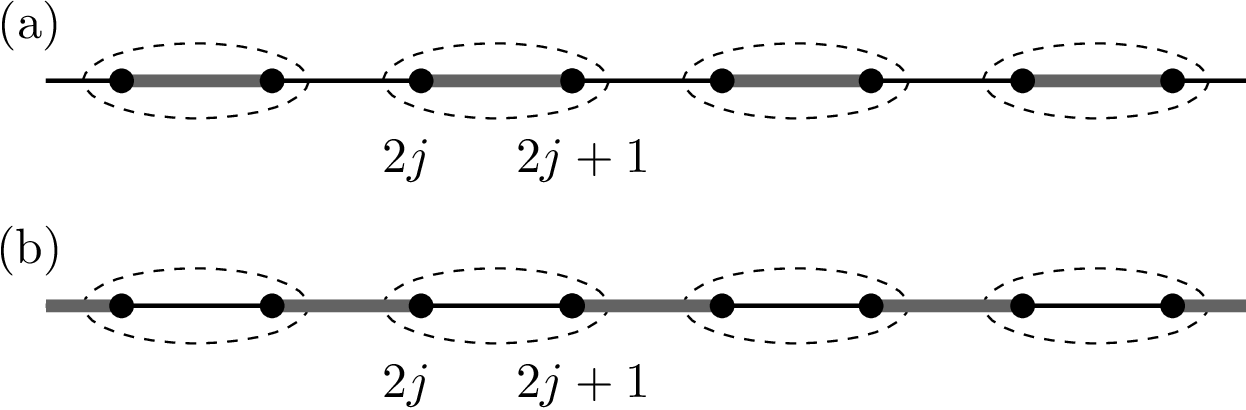,width=10truecm}}
\caption[dummy]{
Schematic pictures of the ground states \rlb{P01}.
(a) depicts the state $\ket{\Psi_0^{\rm SSH}}$, and (b) depicts $\ket{\Psi_1^{\rm SSH}}$.
Here a thick gray line connecting sites $k$ and $k+1$ represent the ``bonding state" created by $(\hcd_{k}-\hcd_{k+1})/\sqrt{2}$.
Note that a bond is confined within a unit cell in the topologically trivial state in (a), while a bond connects two adjacent unit cells in a topologically nontrivial state in (b).

We can also interpret the two diagrams as representing the ground states \rlb{HP01} of the Hubbard model.
In this case, a thick gray line represents a state created by $\hA^\dagger_{k,k+1}$.
}
\label{f:P01}
\end{figure}

\subsection{Simple examples}
An illuminating example is given by the two extreme ground states of the SSH model.
They are written formally as $\omega^{\rm SSH}_s(\cdots)=\bra{\Psi_s^{\rm SSH}}\cdots\ket{\Psi_s^{\rm SSH}}$, where $s=0,1$, with 
\eq
\ket{\Psi_0^{\rm SSH}}=\Bigl(\prod_{j\in\bbZ}\frac{\hcd_{2j}-\hcd_{2j+1}}{\sqrt{2}}\Bigr)\vac,\quad
\ket{\Psi_1^{\rm SSH}}=\Bigl(\prod_{j\in\bbZ}\frac{\hcd_{2j-1}-\hcd_{2j}}{\sqrt{2}}\Bigr)\vac,\lb{P01}
\en
where $\vac$ is the state with no fermions.
See Figure~\ref{f:P01}.
From a direct and elementary computation for a finite periodic chain and Theorem~\ref{t:finitetoinfinite}, we see $\omega^{\rm SSH}_0$ and $\omega^{\rm SSH}_1$ are locally-unique gapped fixed-charge ground states of the Hamiltonians $\hH_0^{\rm SSH}$ and $\hH_1^{\rm SSH}$, respectively,  defined in \rlb{HSSH}.

Note first that $(\hn_{2j}+\hn_{2j+1})\ket{\Psi_0^{\rm SSH}}=\ket{\Psi_0^{\rm SSH}}$ for any $j$.
From this, it is readily found that $\omega_0^{\rm SSH}(\hU_\theta)=1$.
To evaluate $\omega_1^{\rm SSH}(\hU_\theta)$, we write
\eq
\hU_\theta=\exp\bigl[i2\pi(\hn_{2j_1+1}-\tfrac{1}{2})\bigr]\prod_{j=j_0}^{j_1}\hat{u}_j,
\en
with 
\eq
\hat{u}_j=\exp\bigl[i\{\theta(2j-2)(\hn_{2j-1}-\tfrac{1}{2})+\theta(2j)(\hn_{2j}-\tfrac{1}{2})\}\bigr].
\en
Here $j_0$ and $j_1$ are such that $2j_0-2<x_0$ and $2j_1>x_0+\ell-2r_0$.
Since $\exp[i2\pi(\hn_{2j_1+1}-\frac{1}{2})]=-1$ and $\omega_1^{\rm SSH}(\hat{u}_j)=\cos[\{\theta(2j-2)-\theta(2j)\}/2]\simeq1$, we see that  $\omega_1^{\rm SSH}(\hU_\theta)\simeq-1$, for large $\ell$.

Thus our indices for the ground states $\omega_0^{\rm SSH}$ and $\omega_1^{\rm SSH}$ are given by $\Ind_0=1$ and $\Ind_1=-1$, respectively, which are written as $(-1)^\nu$ where $\nu$ is the Zak phase mentioned in section~\ref{s:intro}.\footnote{%
If we use the duality stated in Theorem~\ref{t:duality} below, then $\Ind_0=1$ implies $\Ind_1=-1$.
In this sense, the above (slightly complicated) evaluation of $\omega_1^{\rm SSH}(\hU_\theta)$ is not necessary.
}
To see the implication of this observation, let $\hH_s$ with $s\in[0,1]$ be an arbitrary one-parameter family of short-ranged $\Gphs$-invariant Hamiltonians of the form \rlb{H0}, \rlb{H1} such that $\hH_0=\hH_0^{\rm SSH}$ and $\hH_1=\hH_1^{\rm SSH}$.
Then Corollary~\ref{c:index} implies that the model with $\hH_s$ must undergo a ground state phase transition at an intermediate $s$.
Note that, although $\hH_0=\hH_0^{\rm SSH}$ and $\hH_1=\hH_1^{\rm SSH}$ describe simple non-interacting fermion models, the interpolating $\hH_s$ can be any Hamiltonian (in the present class) with interactions and disorder.
Recall that the Zak phase is defined only for translation-invariant non-interacting models.

\subsection{Duality of the index}
Our definition \rlb{U1} of the twist operator corresponds to our convention to regard sites $2j$ and $2j+1$ as forming a unit cell.
See Figure~\ref{f:chain}.
If we use a different convention, where $2j-1$ and $2j$ form a unit cell, the corresponding twist operator becomes
\eq
\hU'_\theta=\exp\Bigl[i\sum_{j\in\bbZ}\theta(2j)\,(\hn_{2j-1}+\hn_{2j}-1)\Bigr].
\lb{U'}
\en
We then have the corresponding index $\Ind'_\omega\in\{1,-1\}$ for a $\Gphs$-invariant locally-unique gapped fixed-charge ground state $\omega$.
The two indices satisfy the following duality.
\begin{theorem}\label{t:duality}
For any $\Gphs$-invariant locally-unique gapped fixed-charge ground state $\omega$, the two indices satisfy
\eq
\Ind_\omega\times\Ind'_\omega=-1.
\lb{I+I'}
\en
\end{theorem}
We thus see that any $\Gphs$-invariant locally-unique gapped fixed-charge ground state is topologically nontrivial with respect to either $\Ind_\omega$ or $\Ind'_\omega$.
Such a duality is suggested by the symmetry between the two ground states \rlb{P01} of the SSH model.

\subsection{Gapless edge excitations}
A topologically nontrivial ground state defined on a geometry with a boundary is often accompanied by gapless excitations located near the boundary.
Using our index, we can easily prove the existence of such gapless edge modes for models on the half-infinite chain.
There are several different versions of such statements.  Let us present two of them.

Consider the half-infinite chain $\bbZ_+=\{0,1,2,\ldots\}$.
Note that we split the chain so as to preserve the structure of unit cells.
Let
\eq
\hH_+=\sum_{j,k\in\bbZ_+}t_{j,k}\,\hcd_j\hc_k+\sumtwo{j\in\bbZ_+}{(\operatorname{supp}\hv_j\subset\bbZ_+)}\hv_j+\Di\hH,
\en
be the Hamiltonian on $\bbZ_+$ obtained by restringing $\hH$ defined in \rlb{H0} and \rlb{H1} to $\bbZ_+$.
(The second summation is restricted to the range where the support of $\hv_j$ is contained in $\bbZ_+$.)
Here $\Di\hH\in\Aloc$ is a boundary Hamiltonian that acts near the edge and is assumed to conserve particle number and has $\Gphs$-invariance.

To state the first theorem, we assume that the bulk Hamiltonian $\hH$ has a $\Gphs$-invariant locally-unique gapped fixed-charge ground state $\omega$ with nontrivial index $\Ind_\omega=-1$.
We then make a nontrivial assumption that there is a fixed-charge ground state $\omega_+$ of $\hH_+$ that coincides with $\omega$ far from the edge, i.e.,
\eq
\lim_{j\up\infty}\bigl|\omega(\tau_j(\hA))-\omega_+(\tau_j(\hA))\bigr|=0,
\lb{AA}
\en
for any $\hA\in\Aloc$, where $\tau_j$ is the translation by $j\in\bbZ$.
(More precisely $\tau_j$ is a linear $*$-automorphism determined by $\tau_j(\hc_k)=\hc_{k+j}$.)
We then have the following.
\begin{theorem}\label{t:edge1}
For any $\ep>0$, there exists a local unitary operator $\hU_\ep$ that may act nontrivially only on an interval of length $16\pi^2t_0/\ep$, and satisfies $[\hU_\ep,\hN]=0$, $\omega_+(\hU_\ep)=0$, and
\eq
\omega_+(\hU_\ep^\dagger\,[\hH_+,\hU_\ep])\le\ep.
\lb{o+UHU}
\en
\end{theorem}
Note that in a finite quantum system with $\omega_+(\cdot)=\sbkt{\Phi_{\rm GS}^+|\cdot|\Phi_{\rm GS}^+}$ the conditions $\omega_+(\hU_\ep)=0$ and \rlb{o+UHU} read $\sbkt{\Psi|\Phi_{\rm GS}^+}=0$ and $\sbkt{\Psi|\hH_+|\Psi}\le E_{\rm GS}^++\ep$, respectively, where $\hH_+\ket{\Phi_{\rm GS}^+}=E_{\rm GS}^+\ket{\Phi_{\rm GS}^+}$ and $\ket{\Psi}=\hU_\ep\ket{\Phi_{\rm GS}^+}$.
This justifies the interpretation that the theorem establishes the existence of a gapless excitation.

We expect that the interval on which $\hU_\ep$ acts nontrivially includes (or is very close) to the boundary site $j=0$, but there cannot be a generally valid estimate.
In any case, the support of a local operator on the half-infinite chain is always near the edge (compared with the whole chain).
Therefore the existence of a local unitary operator $\hU_\ep$ with the above property shows that $\omega_+$ is accompanied by a particle-number-conserving gapless excitation near the edge.

The above assumption \rlb{AA} about $\omega_+$ is plausible, but we do not know of a general proof of the existence of such a ground state.
Theorem~\ref{t:edge1} nevertheless establishes rigorously that one can avoid gapless edge excitations only in a pathological case where \rlb{AA} is not satisfied.

In the second theorem, we work in the Fock space setting.
We consider a bulk $\Gphs$-invariant Hamiltonian $\hH$ of the form \rlb{H0}, \rlb{H1} that is invariant under translation by a certain even distance $r_1$, i.e., $t_{j,k}=t_{j+r_1,k+r_1}$ and $\tau_{r_1}(\hv_j)=\hv_{j+r_1}$ for any $j,k\in\bbZ$.
We assume that $\hH$ has a unique gapped ground state $\omega$ in the sense of Definition~\ref{d:UGGS}.
(See footnote~\ref{fn:H0}.)
Note that the uniqueness (not the local-uniqueness) implies $\omega$ is $\Gphs$-invariant.
We finally assume that the ground state is topologically nontrivial, i.e., $\Ind_\omega=-1$.
Under these (rather strong) assumptions, we can prove the following strong statement for any ground state of $\hH_+$.
\begin{theorem}\label{t:edge2}
Let $\omega_+$ be an arbitrary ground state of $\hH_+$ in the sense of Definition~\ref{d:GS}.
For any $\ep>0$, there exists a local unitary operator $\hU_\ep$  that may act nontrivially only on an interval of length $16\pi^2t_0/\ep$, and satisfies $[\hU_\ep,\hN]=0$, $\omega_+(\hU_\ep)=0$, and
\eq
\omega_+(\hU_\ep^\dagger\,[\hH_+,\hU_\ep])\le\ep.
\lb{o+UHU2}
\en
\end{theorem}
As discussed above, this means that $\omega_+$ is accompanied by a particle-number-conserving gapless excitation near the edge.

Let us finally comment on an interesting application of the duality stated in Theorem~\ref{t:duality}.
Suppose that the bulk Hamiltonian $\hH$ has a (locally-)unique gapped ground state $\omega$ with trivial index $\Ind_\omega=1$.
Recalling that the same state has nontrivial index $\Ind'_\omega=-1$ according to the other convention of unit cells, we see that the model has gapless edge excitations when restricted onto the half-infinite chain $\{1,2,\ldots\}$.

\subsection{Decoupled systems}
For fixed $j_0\in\bbZ$, let $\hH_{\rm dec}$ be a short-ranged $\Gphs$-invariant Hamiltonian of the form \rlb{H0}, \rlb{H1} with no particle hopping (but possible interactions) between two half-infinite chains $\{\ldots,2j_0-2,2j_0-1\}$ and $\{2j_0,2j_0+1,\ldots\}$.
More precisely, we assume that $t_{j,k}=0$ whenever $j\ge2j_0$, $k<2j_0$ or $j<2j_0$, $k\ge2j_0$.
We then have the following.

\begin{theorem}\label{t:dis}
If $\hH_{\rm dec}$ has a $\Gphs$-invariant locally-unique gapped fixed-charge ground state $\omega_{\rm dec}$, then $\Ind_{\omega_{\rm dec}}=1$, i.e., $\omega_{\rm dec}$ is topologically trivial.
\end{theorem}
We thus see from Corollary~\ref{c:index} that a topologically nontrivial ground state $\omega$ with $\Ind_\omega=-1$ cannot be continuously connected to $\omega_{\rm dec}$ through $\Gphs$-invariant models.
This suggests that $\omega$ has entanglement between the two half-infinite chains that cannot be eliminated by a continuous modification that respects the $\Uone\rtimes\bbZ_2$ symmetry. 
%{\bf We recall that the presence of entanglement protected by symmetry is one of the most essential characteristics of a ground state in nontrivial symmetry-protected topological phases.???}

\section{Proofs}\label{s:proof}
We shall prove all the theorems stated in the previous section.

\subsection{The main index theorem}
To prove our main theorem, Theorem~\ref{t:index}, we state a simple but essential lemma for any Hamiltonian of the form \rlb{H0}, \rlb{H1}.
The lemma is nothing but the standard variational estimate that goes back to Lieb, Schultz, and Mattis \cite{LiebSchultzMattis1961}.
Here the U(1) invariance, as stated in \rlb{U11} and \rlb{U12}, plays an essential role, and the invariance under $\Gphs$ is not assumed.
\begin{lemma}\label{l:LSM}
For any $\theta$-function, one has
\eq
\omega\bigl(\hU_\theta^\dagger[\hH,\hU_\theta]\bigr)\le t_0\gamma^2\ell.
\lb{LSM}
\en
\end{lemma}

\noindent
{\em Proof:}
Recalling that $[\Hint,\hU_\theta]=0$, $\omega(\Gphs(\hA))=\omega(\hA)$, $\Gphs(\Hhop)=\Hhop$, and $\Gphs(\hU_\theta)=\hU_\theta^\dagger$, we see that
\eqa
\omega\bigl(\hU_\theta^\dagger[\hH,\hU_\theta]\bigr)
&=\omega\bigl(\hU_\theta^\dagger[\Hhop,\hU_\theta]\bigr)=\frac{1}{2}\bigr\{\omega\bigl(\hU_\theta^\dagger[\Hhop,\hU_\theta]\bigr)
+\omega\bigl(\Gphs(\hU_\theta^\dagger[\Hhop,\hU_\theta])\bigr)\bigr\}
\nl&=\frac{1}{2}\bigr\{\omega\bigl(\hU_\theta^\dagger[\Hhop,\hU_\theta]\bigr)
+\omega\bigl(\hU_\theta[\Hhop,\hU_\theta^\dagger]\bigr)\bigr\}
=\frac{1}{2}\omega\bigl([\hU_\theta^\dagger,[\Hhop,\hU_\theta]]\bigr)
\nl&\le\frac{1}{2}\Bigl\Vert[\hU_\theta^\dagger,[\Hhop,\hU_\theta]]\Bigr\Vert
\le\frac{1}{2}\sum_{j,k}|t_{j,k}|\,\Bigl\Vert[\hU_\theta^\dagger,[\hcd_j\hc_k,\hU_\theta]]\Bigr\Vert,
\lb{oUHU}
\ena
where we used the expression \rlb{H1} of $\Hhop$.
Let us write \rlb{U1} as $\hU_\theta=\exp[i\sum_j\theta_j(\hn_j-\frac{1}{2})]$ with $\theta_j=\theta(2\lfloor j/2\rfloor)$.\footnote{
In section~\ref{s:other} we use $\hU_\theta$ defined by \rlb{U2} and \rlb{U3}.
In these cases we set $\theta_j=\theta(j+\frac{1}{2})$ and $\theta_j=\theta(j)$, respectively.
}
Noting that $[\hcd_j\hc_k,\hU_\theta]=(e^{i\theta_k}-e^{i\theta_j})\hcd_j\hc_k$ and $[\hU_\theta^\dagger,\hcd_j\hc_k]=(e^{-i\theta_j}-e^{-i\theta_k})\hcd_j\hc_k$, we see that
\eq
[\hU_\theta^\dagger,[\hcd_j\hc_k,\hU_\theta]]=
2\{\cos(\theta_j-\theta_k)-1\}\hcd_j\hc_k.
\lb{UccU}
\en
The norm is readily bounded as 
\eq
\Bigl\Vert[\hU_\theta^\dagger,[\hcd_j\hc_k,\hU_\theta]]\Bigr\Vert\le2\,|\cos(\theta_j-\theta_k)-1|\le(\theta_j-\theta_k)^2
\le\gamma^2(|j-k|+1)^2,
\en
where we used $|\theta'(x)|\le\gamma$.
Then we see from \rlb{oUHU} that 
\eqa
\omega\bigl(\hU_\theta^\dagger[\hH,\hU_\theta]\bigr)&\le\frac{1}{2}\sumtwo{j,k}{(\theta_j\ne\theta_k)}|t_{j,k}|\,\gamma^2(|j-k|+1)^2
\nl
&\le\frac{1}{2}\sum_{j\in[x_0-r_0,x_0+\ell-r_0]\cap\bbZ}\sum_k|t_{j,k}|\,\gamma^2(|j-k|+1)^2\le t_0\gamma^2\ell.\ena
where we used the assumption \rlb{tshort} about the summability of $t_{j,k}$.~\qedm

\medskip
We are now ready to prove Theorem~\ref{t:index}.
Let us assume that $\omega(\hU_\theta)=0$ for some $\theta$-function with $\gamma^2\ell<\DE/t_0$.
Then \rlb{LSM}, along with the definition of the energy gap (see Definition~\ref{d:LUGFCGS}), implies that the gap above the ground state $\omega$ is not greater than $t_0\gamma^2\ell$, which is a contradiction.
We have thus shown that $\omega(\hU_\theta)\ne0$ for any $\theta$-function with $\gamma^2\ell<\DE/t_0$.

To show the independence of the sign, we take a function $\theta$ characterized by $x_0$, $\ell$, and $\gamma$, and another function $\theta'$ characterized by $x'_0$, $\ell'$, $\gamma'$.
We assume $\gamma^2\ell<\DE/t_0$ and $\gamma'^2\ell'<\DE/t_0$.
We claim that there is a family of functions $\theta_s$ that depends continuously on $s\in[0,1]$ such that $\theta_0=\theta$ and $\theta_1=\theta'$.
Moreover  $\theta_s$ is characterized by $\ell_s$ and $\gamma_s$ such that $(\gamma_s)^2\ell_s<\DE/t_0$.
Since the expectation value $\omega(\hU_{\theta_s})\in\bbR$ depends continuously on $s$ and is nonzero, its sign can never change.

The interpolating function $\theta_s$ can be constructed as follows.
We assume $\ell\le\ell'$ without losing generality.
We first rescale the x-coordinate of $\theta$ as $\tilde{\theta}_\kappa(x)=\theta(\kappa(x-x_0)+x_0)$ and vary the scaling factor $\kappa$ from 1 to $(\ell'-2r_0)/(\ell-2r_0)\ge1$.
The function $\tilde{\theta}_\kappa$ is characterized by parameters $\tilde{\gamma}_\kappa=\gamma/\kappa$ and $\tilde{\ell}_\kappa=\kappa\ell-2(\kappa-1)r_0\le\kappa\ell$.
We thus see $(\tilde{\gamma}_\kappa)^2\tilde{\ell}_\kappa\le\gamma^2\ell<\DE/t_0$.
We then translate the rescaled function by $x'_0-x_0$ so that the non-constant part of the resulting function and that of $\theta'$ coincide.
Finally, we linearly interpolate the resulting function and $\theta'$.

\subsection{The duality theorem}
We now prove Theorem~\ref{t:duality}, i.e., the duality $\Ind_\omega\times\Ind'_\omega=-1$.

We assume that the model has a $\Gphs$-invariant locally-unique gapped fixed-charge ground state $\omega$ with energy gap $\DE$.
Take a $\theta$-function \rlb{thetacond} such that $\gamma^2\ell<\DE/t_0$, and consider the two twist operators
\eqg
\hU_\theta=\exp\Bigl[i\sum_{j}\theta(2j)\,(\hn_{2j}+\hn_{2j+1}-1)\Bigr],\lb{Uagain}\\
\hU'_\theta=\exp\Bigl[i\sum_{j}\theta(2j)\,(\hn_{2j-1}+\hn_{2j}-1)\Bigr],
\eng
where \rlb{Uagain} is the same as \rlb{U1}.
The two operators correspond to the two ways of defining unit cells.
Our goal is to prove
\eq
\omega(\hU_\theta)\,\omega(\hU'_\theta)<0,
\lb{goal1}
\en
which implies $\Ind_\omega\times\Ind'_\omega=-1$.
The proof is easy and elementary, but we need to go through two steps.

For the same $\theta$-function, we define
\eq
\hV_\theta=\exp\Bigl[i\sum_{j}\theta(-2j)\,(\hn_{2j}+\hn_{2j+1}-1)\Bigr],
\en
where the phase $\theta(-2j)$ now starts from $2\pi$ and decreases to 0 as $j$ increases.
We shall show
\eq
\omega(\hU_\theta)\,\omega(\hV_\theta)>0.
\lb{goal2}
\en
We let $\theta_R(x)=\theta(x+R)$, and define
\eq
\hW^{(\alpha)}_R=\exp\Bigl[i\alpha\sum_{j<0}\theta_R(2j)\,(\hn_{2j}+\hn_{2j+1}-1)\Bigr]\,
\exp\Bigl[i\alpha\sum_{j\ge0}\theta_R(-2j)\,(\hn_{2j}+\hn_{2j+1}-1)\Bigr],
\en
where $\alpha\in[0,1]$.
Since a gapped ground state is clustering as was established in \cite{NS1,HastingsKoma}, we see that the difference
\eq
\omega(\hW^{(1)}_R)-\omega(\hU_{\theta_R})\,\omega(\hV_{\theta_R})
\en
converges to zero as $R\up\infty$.
(See Remark below.)
On the other hand, since $\hW^{(\alpha)}_R$ is continuous in $\alpha$, and $\omega(\hW^{(0)}_R)=\omega(\iop)=1$, we see (by using the same logic as in the proof of Theorem~\ref{t:index}) that $\omega(\hW^{(1)}_R)>0$.
The two observation together prove $\omega(\hU_{\theta_R})\,\omega(\hV_{\theta_R})>0$ for sufficiently large $R$.
Since the signs of $\omega(\hU_{\theta_R})$ and $\omega(\hV_{\theta_R})$ are independent of $R$ (again by continuity), we have \rlb{goal2}.

We next define
\eq
\hX^{(\alpha)}_R=\exp\Bigl[i\alpha\sum_{j<0}\theta_R(2j)\,(\hn_{2j-1}+\hn_{2j}-1)\Bigr]\,\exp[i\alpha2\pi\,(\hn_{-1}-\tfrac{1}{2})]\,
\exp\Bigl[i\alpha\sum_{j\ge0}\theta_R(-2j)\,(\hn_{2j}+\hn_{2j+1}-1)\Bigr],
\en
with $\alpha\in[0,1]$, where we assume $R\ge x_0+\ell-2r_0$ so as to always have $\theta_R(0)=2\pi$.
Again, by using the clustering property and recalling that $\exp[i2\pi\,(\hn_{-1}-\tfrac{1}{2})]=-1$, we see that
\eq
\omega(\hX^{(1)}_R)\simeq-\omega(\hU'_{\theta_R})\,\omega(\hV_{\theta_R}),
\en
for sufficiently large $R$.
Noting that $\hX^{(0)}_R=\iop$, we see from continuity that
\eq
\omega(\hU'_{\theta})\,\omega(\hV_{\theta})<0,
\en
which, with \rlb{goal2}, implies the desired \rlb{goal1}.

\medskip\noindent
{\em Remark}\/:
To be precise, the conditions for the clustering theorems in \cite{NS1,HastingsKoma} are not satisfied in the present setting.
The spectral gap condition in \cite{NS1} is satisfied for a locally-unique gapped ground state (Definition~\ref{d:LUGGS}) but not necessarily for a locally-unique gapped fixed-charge ground state (Definition~\ref{d:LUGFCGS}).
However when the operators in question commute with $\hN$, then one can apply essentially the same proof as in \cite{NS1} to prove the desired clustering theorem for a locally-unique gapped fixed-charge ground state.
This is sufficient for our purpose.
We thank Alex Bols for clarifying this point.

\subsection{The theorems for gapless edge excitations}
Let us prove Theorem~\ref{t:edge1}.
Let $\DE>0$ be the energy gap of the ground state $\omega$, and fix an arbitrary $\ep$ such that $0<\ep<\DE$.
We take the piecewise linear function $\bar{\theta}_{x_0,\ell}$ defined in \rlb{bartheta}, and set
\eq
\ell=\frac{16\pi^2t_0}{\ep}.
\en
Recall that we then have $\gamma=2\pi/(\ell-2r_0)$.
We shall fix $\ell$ and vary $x_0$ in the following proof.
Assuming that $\ep$ is small enough that we have $\ell\ge r_0/4$ (which is equivalent to $\ell-2r_0\ge\ell/2$), we have
\eq
t_0\gamma^2\ell=t_0\Bigl(\frac{2\pi}{\ell-2r_0}\Bigr)^2\ell\le\frac{16\pi^2t_0}{\ell}=\ep.
\lb{t0ep}
\en

By assumptions we have $\omega(\hU_{\bar{\theta}_{x_0,\ell}})<0$ for any $x_0$.
Note that $\omega_+(\hU_{\bar{\theta}_{x_0,\ell}})-\omega(\hU_{\bar{\theta}_{x_0,\ell}})$ vanishes as $x_0\up\infty$ because of \rlb{AA}.
This means that we have $\omega_+(\hU_{\bar{\theta}_{x_0,\ell}})<0$ for sufficiently large $x_0$.
On the other hand, if $x_0+\ell-2r_0\le0$, we have $\bar{\theta}_{x_0,\ell}(x)=2\pi$ for all $x\ge0$ and hence  $\omega_+(\hU_{\bar{\theta}_{x_0,\ell}})=1$.
By continuity there is $x_0$ at which $\omega_+(\hU_{\bar{\theta}_{x_0,\ell}})=0$.
Let $\hU_{\ep}$ be $\hU_{\bar{\theta}_{x_0,\ell}}$ with this particular $x_0$.
Then \rlb{LSM} and \rlb{t0ep} imply $\omega_+(\hU^\dagger_\ep[\hH_+,\hU_\ep])\le\ep$.

We shall prove Theorem~\ref{t:edge2}.
Consider a state on $\bbZ$ defined by the limit (with possibly taking a subsequence)
\eq
\tilde{\omega}(\hA)=\lim_{n\up\infty}\omega_+(\tau_{nr_1}(\hA)),
\en
where $\tau_j$ denotes the translation by $j$.
One then finds (by essentially repeating the argument in the proof of Theorem~\ref{t:finitetoinfinite}) that $\tilde{\omega}$ is a ground state of $\hH$.
The assumed uniqueness implies that $\tilde{\omega}=\omega$.
Since this justifies the assumption \rlb{AA}, the desired theorem reduces to Theorem~\ref{t:edge1}.

\subsection{The decoupling theorem}
Let us prove Theorem~\ref{t:dis}.
We set $j_0=0$ without losing generality.
Fix a function $\theta$ with  $\gamma^2\ell<\DE/t_0$ (where $\DE$ is the energy gap of $\omega'$) and define a new function $\tilde{\theta}_R$ by
\eq
\tilde{\theta}_R=\begin{cases}
\theta(x-R),&x<0;\\
\theta(x+R),&x\ge0,
\end{cases}
\en
for $R\in\bbR$.
Although $\tilde{\theta}_R(x)$ may be discontinuous at $x=0$, the estimate \rlb{LSM}, with $\theta$ and $\hH$ replaced by $\tilde{\theta}_R$ and $\hH_{\rm dec}$, respectively, is still valid because the relevant hopping is missing.
We thus have
\eq
\omega_{\rm dec}(\hU^\dagger_{\tilde{\theta}_R}[\hH_{\rm dec},\hU_{\tilde{\theta}_R}])<\DE.
\en
As in the proof of Theorem~\ref{t:index}, this implies that $\omega_{\rm dec}(\hU_{\tilde{\theta}_R})\in\bbR\backslash\{0\}$ for any $R$, and, hence, by continuity, that $\omega_{\rm dec}(\hU_{\tilde{\theta}_R})$ has a well-defined sign independent of $R$.

By letting $R$ sufficiently large, we have $\tilde{\theta}_R(x)=2\pi$ for $x\ge0$ and $\tilde{\theta}_R(x)=0$ for $x<0$, and hence $\omega_{\rm dec}(\hU_{\tilde{\theta}_R})=1$.
Noting that $\tilde{\theta}_0$ is nothing but the original $\theta$, we see $\omega_{\rm dec}(\hU_\theta)>0$, which means $\Ind_{\omega_{\rm dex}}=1$.

\section{Other models}\label{s:other}
Let us briefly discuss some classes of models in which we can develop rigorous index theory by using the same techniques as in the SSH-type models.

\subsection{Models of electrons}
We discuss an important extension to models of electrons, such as the Hubbard model, on the infinite chain.
We here do not attempt at presenting general classes of models but concentrate on one typical class.

This extension is interesting as it is directly connected to the problem of the Haldane phase, i.e., the nontrivial symmetry-protected topological (SPT) phase, in $S=1$ quantum spin chains.
See, e.g., Part~2 of \cite{TasakiBook} for details.
See also \cite{Zirnbauer,Sompetetal} for the relation between the extended Hubbard model and $S=1$ quantum spin chains.

\paragraph{General models and the index theory}
Let $\hcd_{j,\sigma}$ and $\hc_{j,\sigma}$ be the standard creation and annihilation operators, respectively, of an electron at site $j\in\bbZ$ with spin $\sigma\in\uds$.
They satisfy the canonical anticommutation relations 
\eq
\{\hc_{j,\sigma},\hc_{k,\tau}\}=0,\quad\{\hc_{j,\sigma},\hcd_{k,\tau}\}=\delta_{j,k}\,\delta_{\sigma,\tau},
\lb{CAC2}
\en
for any $j,k\in\bbZ$ and $\sigma,\tau\in\uds$.
We again write the Hamiltonian as $\hH=\Hhop+\Hint$ with
\eq
\Hhop=\sumtwo{j,k\in\bbZ}{\sigma\in\uds}t_{j,k}\,\hcd_{j,\sigma}\hc_{k,\sigma},\quad
\Hint=\sum_{j\in\bbZ}\hv_j,
\lb{H2}
\en
where $t_{j,k}=(t_{k,j})^*\in\bbC$ and $\hv_j^\dagger=\hv_j$.
In the following, we only consider examples with $t_{j,k}=t_{k,j}\in\bbR$.
We assume the same short-range conditions for $t_{j,k}$ and $\hv_j$ as in section~\ref{s:U1}.

The Hamiltonian $\hH$ of a reasonable model of electrons (except for effective models for superconductivity) conserves the total particle number $\hN$.
Here we make a stronger assumption that the number of up-spin electrons and down-spin electrons, namely, $\hN_{\up}:=\sum_{j\in\bbZ}\hn_{j,\up}$ and $\hN_{\dn}:=\sum_{j\in\bbZ}\hn_{j,\dn}$, are conserved separately.
Since it holds that $[t_{j,k}\,\hcd_{j,\sigma}\hc_{k,\sigma},\hN_\tau]=0$ for any $\sigma,\tau\in\uds$, our nontrivial assumption is that $[\hv_j,\hN_{\up}]=[\hv_j,\hN_{\dn}]=0$ for any $j\in\bbZ$.
(Here the commutators are interpreted as in \rlb{AN}.)
This means that $\hv_j$ can be a polynomial of $\hn_{k,\sigma}$ or something slightly more complicated like the Heisenberg exchange interaction
\eq
-J\,\hbS_{j}\cdot\hbS_k=-J\Bigl\{\frac{1}{2}(\hcd_{j,\up}\hc_{j,\dn}\hcd_{k,\dn}\hc_{k,\up}+\text{h.c.})+
\frac{1}{4}(n_{j,\up}-n_{j,\dn})(n_{k,\up}-n_{k,\dn})\Bigr\},
\lb{HEI}
\en
where $J$ is a real constant, and $\hbS_j=(\hSx_j,\hSy_j\hSz_j)$ denote the spin operators at site $j$.\footnote{%
The spin operators are defined in terms of the fermion operators as
$\hSz_j=(\hn_{j,\up}-\hn_{j,\dn})/2$, $\hat{S}^+_j=\hSx_j+i\hSy_j=\hcd_{j,\up}\hc_{j,\dn}$, and $\hat{S}^-_j=\hSx_j-i\hSy_j=\hcd_{j,\dn}\hc_{j,\up}$.
}
We note that the assumed symmetry corresponds to (or leads to) the U(1) symmetry with respect to an arbitrary uniform spin rotation about the z-axis in quantum spin systems.

We need an additional $\bbZ_2$ symmetry to guarantee the reality of the expectation value $\omega(\hU_\theta)$.
There are some choices (see section~\ref{s:othersymmetry} below) but let us concentrate on the most important example, namely, the combination of the particle-hole transformation and the gauge transformation.
This is the same symmetry as in the SSH-type models studied in the previous sections.
It is described by the linear $*$-automorphism $\Gphs$ defined by
\eq
\Gphs(\hc_{j,\sigma})=(-1)^j\,\hcd_{j,\sigma},
\en
for any $j\in\bbZ$ and $\sigma\in\uds$.
Note that any $\Gphs$-invariant state $\rho$ satisfies $\rho(\hn_{j,\sigma})=1/2$ for any $j$ and $\sigma$, i.e., it describes the system at half-filling.

We then follow \rlb{U1} and define the twist operator by
\eq
\hU_\theta=\exp\Bigl[i\sum_{j}\theta(2j)\,(\hn_{2j,\up}+\hn_{2j+1,\up}-1)\Bigr],
\lb{Ue}
\en
with the same $\theta$-function \rlb{thetacond}.
It is crucial to note that we only twist the phase of up-spin electrons.
This is the reason we assumed that $\hN_\up$ and $\hN_\dn$ are conserved separately.

Then the rest is exactly the same as in sections~\ref{s:SSH}, \ref{s:Indextheory}, and \ref{s:proof}.
If $\omega$ is a $\Gphs$-invariant locally-unique gapped fixed-charge\footnote{%
To characterize fixed-charge ground states, we use perturbations $\hV\in\Aloc$ such that $[\hV,\hN_\up]=[\hV,\hN_\dn]=0$.
Note that we are making a slight abuse of notation here since not only the total charge but also the total $\hat{S}^{\rm z}$ is conserved.
} ground state of $\hH$, then one finds, as in Theorem~\ref{t:index}, that $\omega(\hU_\theta)\ne0$ and $\omega(\hU_\theta)\in\bbR$ for suitable $\theta$-function.
We can define the index $\Ind_\omega\in\{1,-1\}$ exactly as in Definition~\ref{d:index}.
Other results, namely, Corollary~\ref{c:index}, Theorems~\ref{t:duality}, \ref{t:edge1}, \ref{t:edge2}, and \ref{t:dis} are proved in exactly the same manner.

Let us see two concrete examples.

\paragraph{Example: the SSH-type Hubbard model}
A typical and instructive example is the Hubbard model with the SSH-type hopping amplitude with the Hamiltonian
\eq
\hH_{\rm Hub}=\sumtwo{j\in\bbZ}{\sigma\in\uds}
\bigl\{(1-s)(\hcd_{2j,\sigma}\hc_{2j+1,\sigma}+\text{h.c.})+s(\hcd_{2j-1,\sigma}\hc_{2j,\sigma}+\text{h.c.})\bigr\}
+U\sum_{k\in\bbZ}\hn_{k,\up}\hn_{k,\dn},
\lb{HubSSH}
\en
where $s\in[0,1]$ and $U\in\bbR$ are model parameters.

Again the ground state can be computed explicitly in the two extreme cases with $s=0$ and 1.
By an elementary calculation for the Hubbard model with two sites (see, e.g., Problem~10.1.a (p.~342) of \cite{TasakiBook}), one sees that for any $U\in\bbR$ the unique ground states (at half-filling) for $s=0$ and 1 are again  written formally as $\omega_s(\cdots)=\bra{\Psi_s}\cdots\ket{\Psi_s}$ with 
\eq
\ket{\Psi_0}=\Bigl(\prod_{j\in\bbZ}\hA^\dagger_{2j,2j+1}\Bigr)\vac,\quad
\ket{\Psi_1}=\Bigl(\prod_{j\in\bbZ}\hA^\dagger_{2j-1,2j}\Bigr)\vac,
\lb{HP01}
\en
respectively.
See Figure~\ref{f:P01}.
Here we defined 
\eq
\hA^\dagger_{j,k}:=a_0\Bigl\{4(\hcd_{j,\up}\hcd_{k,\dn}-\hcd_{j,\dn}\hcd_{k,\up})+\bigl(U-\sqrt{U^2+16}\bigr)(\hcd_{j,\up}\hcd_{j,\dn}+\hcd_{k,\up}\hcd_{k,\dn})\Bigr\},
\en
where $a_0=1/\sqrt{32+2(U-\sqrt{U^2+16})^2}$ is the normalization constant.
Note that $\hA^\dagger_{j,k}$ creates a spin-singlet state on two sites $j$ and $k$.

We stress that the nature of the state created by $\hA^\dagger_{j,k}$  depends crucially on $U$.
When $U=0$,
\eq
\hA^\dagger_{j,k}=-\frac{1}{2}(\hcd_{j,\up}-\hcd_{k,\up})(\hcd_{j,\dn}-\hcd_{k,\dn}),
\en
describes two non-interacting electrons with $\up$ and $\dn$ spins occupying the ``bonding'' state $(\hcd_{j}-\hcd_{k})/\sqrt{2}$ as in \rlb{P01} for the SSH model.
In the limit $U\up\infty$ with infinitely large repulsive interaction, we have
\eq
\hA^\dagger_{j,k}=\frac{1}{\sqrt{2}}(\hcd_{j,\up}\hcd_{k,\dn}-\hcd_{j,\dn}\hcd_{k,\up}),
\en
which generates the spin-singlet state on $j$ and $k$ of the effective quantum spin system where each site is occupied by exactly one electron.
Finally, in the limit $U\dn-\infty$, we see
\eq
\hA^\dagger_{j,k}=-\frac{1}{\sqrt{2}}(\hcd_{j,\up}\hcd_{j,\dn}+\hcd_{k,\up}\hcd_{k,\dn}),
\en
which describes two electrons forming on-site spin-singlets because of the infinitely large attractive interaction.

Recall, however, that our twist operator \rlb{Ue} depends only on up-spin electrons.
If we only concentrate on the up-spin electron, $\hA^\dagger_{j,k}$ always creates an equal-weight superposition of two states where the up-spin electron is on $j$ and on $k$.
This means that the evaluation of the expectation values $\omega_0(\hU_\theta)$ and $\omega_1(\hU_\theta)$, and hence that of the indices, becomes exactly the same as in the ground states \rlb{P01} of the SSH model.
We see $\Ind_0=1$ and $\Ind_1=-1$ for the ground states \rlb{HP01} with any $U\in\bbR$.
We thus found rigorously that the ground state $\omega_s$ of \rlb{HubSSH} is in the trivial phase when $s=0$ and is in the nontrivial phase when $s=1$, and hence undergoes a topological phase transition at an intermediate $s\in(0,1)$.
Interestingly, $\omega_1$ describes a topologically nontrivial state even in the limit $U\dn-\infty$, where the state is entirely trivial from the point of view of quantum spin systems.

\paragraph{Example: the extended Hubbard model}
Another interesting example is the extended Hubbard model with the Hamiltonian
\eq
\hH_{\rm ex}=\sumtwo{j\in\bbZ}{\sigma\in\uds}(\hcd_{2j-1,\sigma}\hc_{2j,\sigma}+\text{h.c.})
+U\sum_{k\in\bbZ}\hn_{k,\up}\hn_{k,\dn}
-J\sum_{j\in\bbZ}\hbS_{2j}\cdot\hbS_{2j+1},
\lb{exHub}
\en
where the exchange interaction is given in \rlb{HEI}.
Note that we have introduced the ferromagnetic interaction (Hund's coupling) for electrons in a unit cell.
See Figure~\ref{f:chain}.
For $J=0$, the Hamiltonian $\hH_{\rm ex}$ is the same as $\hH_{\rm Hub}$ of \rlb{HubSSH} with $s=1$, and hence we know that its ground state has $\Ind=-1$ and belongs to the nontrivial topological phase.
For $U\gg1$ and $J\gg1$, it is expected that the low-energy properties of the model \rlb{exHub} at half-filling coincide with those of the $S=1$ Heisenberg antiferromagnetic chain with the Hamiltonian
\eq
\hH_{\rm Heis}=\sum_{k\in\bbZ}\hbS_{k}\cdot\hbS_{k+1},
\lb{S=1H}
\en
where two sites $2j$ and $2j+1$ in the electron model correspond to a single spin at site $k$.
(In \rlb{S=1H}, $\hbS_{k}$ denotes the spin operator with $S=1$.)
It has been conjectured that $\hH_{\rm Heis}$ has a unique gapped ground state with nontrivial index $\Ind=-1$.
The corresponding nontrivial topological phase (or, more precisely, nontrivial symmetry-protected topological phase) is known as the Haldane phase.
See, e.g., Part~2 of \cite{TasakiBook}.
This suggests that the ground state of \rlb{exHub} does not undergo a phase transition when $J$ is increased from 0, but it seems to be extremely difficult to justify this expectation rigorously.

\subsection{Fermion models with different symmetries}
\label{s:othersymmetry}
Let us again focus on general models of spinless fermions with U(1) symmetry defined in section~\ref{s:U1}.

Our index theory for the SSH-type models relies on the U(1) symmetry, i.e., the conservation of the total particle number $\hN$, and the extra $\bbZ_2$ symmetry generated by $\Gphs$ discussed in section~\ref{s:Z2}.
The U(1) symmetry is essential for the validity of the basic variational estimate in Lemma~\ref{l:LSM}, and the $\bbZ_2$ symmetry guarantees the reality of the expectation value $\omega(\hU_\theta)$.
Although the class of models with $\Gphs$ symmetry is natural (and includes the SSH model), that is not the only possibility.
Here we shall briefly discuss other classes of models with different $\bbZ_2$ symmetry.

\paragraph{Particle-hole symmetry}
The particle-hole transformation is described by the liner $*$-automorphism $\Gph$ defined by $\Gph(\hc_j)=\hcd_j$ for any $j$.
The condition that the Hamiltonian \rlb{H0}, \rlb{H1} is invariant under $\Gph$ is that $t_{j,k}$ is pure imaginary for any $j,k$ and $t_{j,j}=0$, and that $\hv_j$ is an even polynomial of $(n_k-\frac{1}{2})$.

Since the twist operator \rlb{U1} satisfies $\Gph(\hU_\theta)=\hU_\theta^\dagger$, one can replace $\Gphs$ with the $\Gph$, and repeat all the discussion in sections~\ref{s:Indextheory} and \ref{s:proof}.
We can define index $\Ind_\omega\in\{1,-1\}$ for a $\Gph$-invariant locally-unique gapped fixed-charge ground state $\omega$ by Theorem~\ref{t:index} and Definition~\ref{d:index}, and then prove Corollary~\ref{c:index}, Theorems~\ref{t:duality}, \ref{t:edge1}, \ref{t:edge2}, and \ref{t:dis} in exactly the same manner.

\paragraph{Bond-centered inversion symmetry}
Let us also discuss two classes of models with inversion symmetry.
In these models, our index may not be sufficient to fully characterize the topological phases, which are believed to be classified by  $\bbZ_4$. See, e.g., \cite{Kapustin}.

The bond-centered inversion transformation $\Gbi$ is defined by $\Gbi(\hc_j)=\hc_{-1-j}$.
We assume that $\hH$ is invariant under $\Gbi$.
For the Hamiltonian \rlb{H0}, \rlb{H1} the conditions for invariance are $t_{j,k}=t_{-1-j,-1-k}$ for any $j,k\in\bbZ$ and $\Gbi(\hv_j)=\hv_{-1-j}$ for any $j\in\bbZ$.
We assume that $\hH$ has a $\Gbi$-invariant locally-unique gapped fixed-charge ground state $\omega$.
Unlike in the SSH-type models, the symmetry does not fix the filling factor of the ground state.
This means that the chemical potential $\mu$ in the Hamiltonian \rlb{Hmu} for the Fock space setting may not be zero.

Here we require that the $\theta$-function is antisymmetric with respect to inversion as
\eq
\theta(x)=2\pi-\theta(-x),
\lb{tsym}
\en
and define the twist operator by
\eq
\hU_\theta=\exp\Bigl[i\sum_{j}\theta(j+\tfrac{1}{2})\,\hn_j\Bigr].
\lb{U2}
\en
Note that this is essentially different from \rlb{U1}. We do not have the sublattice structure here.
Since \rlb{tsym} implies $\Gbi(\hU_\theta)=\hU_\theta^\dagger$, we get the crucial reality condition  $\omega(\hU_\theta)\in\bbR$ from the invariance.
Then the rest is the same, and we can define index by Theorem~\ref{t:index} and Definition~\ref{d:index}, and prove Corollary~\ref{c:index}.
The two ground states \rlb{P01} of the SSH model provide good examples, where we again have $\Ind_{\omega_0}=1$ and $\Ind_{\omega_1}=-1$.
Theorem~\ref{t:dis} with $j_0=0$ is also valid for this class of models, but other theorems do not extend to this class.

\paragraph{Site-centered inversion symmetry}
Models with site-centered inversion symmetry can be treated in almost the same manner by replacing $\Gbi$ with $\Gsi$ defined by $\Gsi(\hc_j)=\hc_{-j}$.
We still assume the symmetry \rlb{tsym} and define
\eq
\hU_\theta=\exp\Bigl[i\sum_{j}\theta(j)\,\hn_j\Bigr].
\lb{U3}
\en
Then we have Theorem~\ref{t:index}, Definition~\ref{d:index}, and Corollary~\ref{c:index}.
Other theorems do not extend to this class.

As an example, consider two atomic states
\eq
\ket{\Psi_{\rm even}}=\Bigl(\prod_j\hcd_{2j}\Bigr)\vac,\ 
\ket{\Psi_{\rm odd}}=\Bigl(\prod_j\hcd_{2j+1}\Bigr)\vac,
\lb{Po}
\en
which are ground states of simple models with alternating on-site potential.
It is easily found that these states have indices $\Ind_{\rm even}=-1$ and $\Ind_{\rm odd}=1$, and hence one inevitably encounters a topological phase transition when the two trivial models are interpolated by a path of models with site-centered inversion symmetry.
See \cite{Fuji} for a similar observation in quantum spin chains.

\subsection{Models of bosons}
We can also treat a system of bosons on the infinite chain.
One of the essential differences from fermion models is that the number of particles on each site is no longer bounded.
We do not treat hardcore boson systems since they are equivalent to quantum spin systems, where rigorous index theorems have already been developed \cite{Tasaki2018,TasakiBook,Ogata1,Ogata2}.

We again consider the Hamiltonian given by \rlb{H0} and \rlb{H1}, but now $\hcd_j$ and $\hc_j$ are the creation and annihilation operators, respectively, of a boson at site $j\in\bbZ$.
They satisfy the commutation relation $[\hc_j,\hcd_k]=\delta_{j,k}$ for any $j,k\in\bbZ$, and are unbounded.
Here we consider classes of models with bond-centered or site-centered inversion symmetry defined exactly as in fermion models.
Interestingly one can repeat the discussion and proofs for fermion models except for one point (see below) and derive the same results, which we do not repeat.
The ground states \rlb{Po} can also be interpreted as those of boson models.

The only point that is different from fermions is that the operator $\hcd_j\hc_k$ that appears on the right-hand side of \rlb{UccU} is unbounded.
Instead of the simple norm bound, we here use the Schwarz inequality as $|\omega(\hcd_j\hc_k)|\le\sqrt{\omega(\hn_j)\omega(\hn_k)}\le\rho_0$, where we made an additional assumption that $\omega(\hn_j)\le\rho_0$ with a constant $\rho_0$ for any $j$.
We then have Lemma~\ref{l:LSM} with the right-hand side of \rlb{UccU} replaced with $\rho_0t_0\gamma^2\ell$.

\section{Discussion}
We have presented a rigorous but elementary index theory for some classes of one-dimensional interacting topological insulators.
For a $\Uone\rtimes\bbZ_2$ invariant locally-unique gapped (fixed-charge) ground state, we defined a $\bbZ_2$ index as the sign of the expectation value of the U(1) twist operator as in Theorem~\ref{t:index} and Definition~\ref{d:index}.
The invariance of the index under continuous modification stated in Corollary~\ref{c:index} is indeed a straightforward consequence of the definition.
For the SSH-type models, which belong to class D in the standard classification, the $\bbZ_2$ topological index is expected to give a complete classification of the (symmetry-protected) topological phases.
We then proved the results concerning the gapless edge modes in Theorems~\ref{t:edge1} and \ref{t:edge2}, and the interesting duality of the index, Theorem~\ref{t:duality}, which reflects the choice of unit cells in the infinite chain.

A core of our index theory is Lemma~\ref{l:LSM}, which goes back to the classical work by Lieb, Schultz, and Mattis \cite{LiebSchultzMattis1961}.
It states that, when acting on a U(1) invariant state, the twist operator $\hU_\theta$ changes the energy expectation value not more than $t_0\gamma^2\ell$.
In the original works of Lieb, Schultz, and Mattis \cite{LiebSchultzMattis1961} and Affleck and Lieb \cite{AffleckLieb1986}, the lemma was applied to a ground state $\omega$ with the condition $\omega(\hU_\theta)=0$, which means that the ground state and its perturbation by $\hU_\theta$ are orthogonal.
Since  $t_0\gamma^2\ell$ can be made as small as one wishes, this shows that the ground state {\em cannot}\/ be locally-unique and gapped, which is the statement of the celebrated Lieb-Schultz-Mattis theorem.
It is interesting that in Theorem~\ref{t:index}, the starting point of our index theory, the same lemma is used in the opposite manner.
We start from the assumption that the ground state $\omega$ is locally-unique and gapped, and then conclude that $\omega(\hU_\theta)\ne0$.
The semi-constructive proof of Theorems~\ref{t:edge1} and \ref{t:edge2} on the existence of a gapless edge mode, on the other hand, follows the original philosophy.
We use the twist operator $\hU_\theta$ to construct a low energy state orthogonal to the ground state as in \cite{LiebSchultzMattis1961,AffleckLieb1986}.
Here the orthogonality condition $\omega(\hU_\theta)=0$ is guaranteed not by the translation symmetry as in \cite{LiebSchultzMattis1961,AffleckLieb1986} but by the continuity of the expectation value.

We believe that our strategy to focus on the expectation value of a {\em local}\/ operator in the {\em infinite}\/ chain makes our theory conceptually simple.
Although we have treated full-fledged infinite quantum systems, our arguments are essentially that of physicists working on a finite system (except that we have the freedom to make $\ell$ as large as we wish).
Recall that, in most rigorous index theories for infinite systems \cite{AvronSeiler,BSB,Matsui2020,BourneOgata,Ogata4,BachmannNachtergaele2014A,Ogata1,Ogata2,Ogata3}, one must deal with sophisticated operator algebraic objects whose connection to the physical picture is sometimes hard to understand.
One can develop a rigorous index theory starting from a finite system as in \cite{Bachmann1,Bachmann2}, but then one has to control the infinite volume limit by a sophisticated argument.

The conceptual simplicity of our proof motivates us to extend our strategy to a larger class of interacting models with topological phases.
An important question is whether the present approach can be extended to higher dimensions.
Clearly, the whole theory readily extends to models defined on the infinite lattice $\bbZ\times\Lambda$, where $\Lambda$ is an arbitrary finite lattice with an odd number of sites.
The twist operator should be defined as $\hU_\theta=\exp[i\sum_{j\in\bbZ,\,q\in\Lambda}\theta(j)\,\hn_{j,q}]$ instead of \rlb{U3}, for example, with the same $\theta(x)$ as before.
We can then prove that, for each arbitrary but fixed $\Lambda$, any locally-unique gapped fixed-charge ground state (with suitable symmetry) is characterized by a $\bbZ_2$-valued topological index.
This establishes the existence of a phase transition in any continuous path of models (with suitable conditions), again with a fixed $\Lambda$.
Whether one can construct a full-fledged index theorem for higher dimensions based on this observation and insights from previous studies \cite{Bachmann1,Bachmann2,Ogata4,WatanabeOshikawa,Hatsugai2006,ShiozakiShapourianRyu,Lu,Matsugatani,Ono,Kang,Wheeler,Nakamura} is an intriguing question.

It is also interesting to extend the present results to some classes of non-Hermitian models that preserve particle number and have suitable symmetry.

\medskip
Finally, let us briefly comment on integer-valued topological indices defined in terms of the expectation value of the twist operator.

Consider a general class of short-ranged Hamiltonians \rlb{H0}, \rlb{H1} without making any assumptions about symmetry except for the (built-in) U(1) symmetry.
We then take a continuous closed path $\hH_s$, with $s\in[0,1]$ and $\hH_0=\hH_1$, of Hamiltonians, and assume that $\hH_s$ has a locally-unique gapped fixed-charge ground state $\omega_s$ (with a uniformly nonzero gap) and that $\omega_s$ depends continuously on $s$.
We consider the simplest twist operator as in \rlb{U3} with a fixed $\theta$-function with sufficiently small $\gamma^2\ell$.
Then the expectation value $\omega_s(\hU_\theta)$ is nonzero and depends continuously on $s$.
Since $\omega_s(\hU_\theta)$ is not necessarily real in this general setting, we see that the path of Hamiltonians determines a closed path in $\bbC\backslash\{0\}$ and hence the winding number $q\in\bbZ$ about the origin.

It is easily found (as in the proof of Corollary~\ref{c:index}) that the winding number $q$ is invariant under any continuous modifications of the paths of Hamiltonians.
In fact, the index $q$ reduces to the Chern number (see, e.g., \cite{AOP}) for non-interacting models with translation invariance.
Recall that the Chern number is precisely the amount of charge transferred in the process of Thouless pumping \cite{Thouless,AOP}.
It is expected that $q$ generally coincides with the number of particles pumped in the cyclic process given by $\hH_s$.
See \cite{TasakiNext} for a partial justification of this expectation and also a relation between spin pumping and SPT phases.

We remark that one may also define a Chern-number-like $\bbZ$-valued index for a given U(1) invariant locally-unique gapped (fixed-charge) ground state $\omega$.
For a one-parameter continuous family $\theta_s$ with $s\in[0,1]$ of $\theta$-functions (with sufficiently small $\gamma^2\ell$) such that $\omega(\hU_{\theta_0})=\omega(\hU_{\theta_1})$, one has a well defined winding number of the continuous path in $\bbC\backslash\{0\}$ determined by $\omega(\hU_{\theta_s})$.
An important example is given by a ground state invariant under translation by $r_1$ and $\theta_s(x)=\theta(x-s r_1)$, in which case the index is nothing but the filling factor.
This fact was used in \cite{TasakiLSM} to prove the generalized Lieb-Schultz-Mattis theorem.

\section*{Acknowledgement}
It is a pleasure to thank Ken Shiozaki for indispensable discussion and comments, and Sven Bachmann, Alex Bols, Chris Bourne, Yoshiko Ogata, Masaki Oshikawa, Yasuhiro Hatsugai,  and Haruki Watanabe for useful comments and discussions.
The present work was supported by JSPS Grants-in-Aid for Scientific Research No. 22K03474.

%%%%%%%%%%%%%%%%%%%%%%%%


\begin{thebibliography}{10}


\bibitem{video:short}
H. Tasaki, {\em Rigorous Index Theory for One-Dimensional Interacting Topological Insulators: a Brief Introduction}\/,
(online lecture (21:41), 2021).
\\\url{https://www.gakushuin.ac.jp/~881791/OL/#Index1DTI2021S}
\\\url{https://youtu.be/ypGVb3eYrpg}

\bibitem{video:long}
H. Tasaki, {\em Rigorous Index Theory for One-Dimensional Interacting Topological Insulators: with a Pedagogical Introduction to the Topological Phase Transition in the SSH Model}\/,
(online lecture (49:07), 2021).
\\\url{https://www.gakushuin.ac.jp/~881791/OL/#Index1DTI2021L}
\\\url{https://youtu.be/_yxZYOevV2Y}

\bibitem{HasnKane}
M. Hasan and C.L. Kane, {\em Colloquium: Topological insulators}\/, Rev. Mod. Phys. {\bf 82}, 3045--3067 (2010).
\\\url{https://arxiv.org/abs/1002.3895}

\bibitem{Ryu}
S. Ryu, A.P. Schnyder, A. Furusaki, and A.W.W. Ludwig, {\em Topological insulators and superconductors: Tenfold way
and dimensional hierarchy}\/, New J. Phys. {\bf 12}, 065010 (2010).
\\\url{https://arxiv.org/abs/0912.2157v2}

\bibitem{Kitaev}
A.Y. Kitaev, {\em Periodic table for topological insulators and superconductors}\/, AIP Conf. Proc. {\bf 1134}, 22--30 (2009).
\\\url{https://arxiv.org/abs/0901.2686}

\bibitem{Shankar}
R. Shankar,
{\em Topological Insulators --- A review}\/, (unpublished note, 2018).
\\\url{https://arxiv.org/abs/1804.06471}

\bibitem{PSB}
E. Prodan, H. Schulz-Baldes, 
{\em Bulk and Boundary Invariants for Complex Topological Insulators: From K-Theory to Physics}\/, 
(Springer, 2016).
\\\url{https://arxiv.org/abs/1510.08744}

\bibitem{KatsuraKoma}
H. Katsura and T. Koma,
{\em The noncommutative index theorem and the periodic table for disordered topological insulators and superconductors}\/,
J. Math. Phys. {\bf 59}, 031903 (2018).
\\\url{https://arxiv.org/abs/1611.01928}





\bibitem{Rachel}
S. Rachel,
{\em Interacting topological insulators: a review}\/,
Rep. Prog. Phys. {\bf 81}, 116501 (2018).
\\\url{https://arxiv.org/abs/1804.10656}







\bibitem{Hatsugai2006}%Berry phase
Y. Hatsugai,
{\em Quantized Berry Phases as a Local Order Parameter of a Quantum Liquid}\/,
J. Phys. Soc. Jpn. {\bf 75}, 123601 (2006).
\\\url{https://journals.jps.jp/doi/pdf/10.1143/JPSJ.75.123601}



\bibitem{GuoShen}%Berry phase
H. Guo and S.-Q. Shen,
{\em Topological phase in a one-dimensional interacting fermion system}\/,
Phys. Rev. B {\bf 84}, 195107 (2011).
\\\url{https://arxiv.org/abs/1108.4996}

\bibitem{FK}
L. Fidkowski and A. Kitaev, {\em Topological phases of fermions in one dimension}\/, 
Phys. Rev. B {\bf 83}, 075103  (2011).
\\\url{https://arxiv.org/abs/1008.4138}


\bibitem{ShiozakiShapourianRyu}
K. Shiozaki, H. Shapourian, and S. Ryu,
{\em Many-body topological invariants in fermionic symmetry protected topological phases: Cases of point group symmetries}\/,
Phys. Rev. B {\bf 95}, 205139 (2017).
\\\url{https://arxiv.org/abs/1609.05970}


\bibitem{Manmana}%Greenfunction
S.R. Manmana,  A.M. Essin, R.M. Noack, and V. Gurarie
{\em Topological invariants and interacting one-dimensional fermionic systems}\/,
Phys. Rev. B {\bf 86}, 205119 (2012).
\\\url{https://arxiv.org/abs/1205.5095}

\bibitem{WangXuWangWu2015}%entanglement entropy
D. Wang, S. Xu, Y. Wang, and C. Wu,
{\em Detecting edge degeneracy in interacting topological insulators through entanglement entropy}\/,
Phys. Rev. B {\bf 91}, 115118 (2015).
\\\url{https://arxiv.org/abs/1405.2043}


\bibitem{Stehouwer}
L. Stehouwer,
{\em Interacting SPT phases are not Morita invariant}\/, 
Lett. Math. Phys. {\bf 112}, 64 (2022).
\\\url{https://arxiv.org/abs/2110.07408v1}


\bibitem{Kapustin}
A. Kapustin, R. Thorngren, A. Turzillo and Z. Wang,
{\em Fermionic symmetry protected topological phases and cobordisms}\/,
J.  High Energy Phys. {\bf 2015}, 1--21 (2015).
\\\url{https://link.springer.com/article/10.1007/JHEP12(2015)052}

%%% higher dim
\bibitem{Lu}
Y.-M. Lu, Y. Ran, and M. Oshikawa,
{\em Filling-enforced constraint on the quantized Hall conductivity on a periodic lattice}\/,
Ann. Phys. {\bf 413}, 168060 (2020).
\\\url{https://arxiv.org/abs/1705.09298}

\bibitem{Matsugatani}
A. Matsugatani, Y. Ishiguro, K. Shiozaki, and H. Watanabe,
{\em Universal Relation among Many-Body Chern Number, Rotation Symmetry, and Filling}\/,
Phys. Rev. Lett. {\bf 120}, 096601 (2018).
\\\url{https://arxiv.org/abs/1710.07012}

\bibitem{Ono}
S. Ono, L. Trifunovic, and H. Watanabe,
{\em Difficulties in operator-based formulation of the bulk quadrupole moment}\/,
Phys. Rev. B {\bf 100}, 245133 (2019).
\\\url{https://arxiv.org/abs/1902.07508}



\bibitem{Kang}
B. Kang, K. Shiozaki, and G.Y. Cho,
{\em Many-body order parameters for multipoles in solids}\/,
Phys. Rev. B {\bf 100}, 245134 (2019).
\\\url{https://arxiv.org/abs/1812.06999}

\bibitem{Wheeler}
W.A. Wheeler, L.K. Wagner, and T.L. Hughes,
{\em Many-body electric multipole operators in extended systems}\/,
Phys. Rev. B {\bf 100}, 245135 (2019).
\\\url{https://arxiv.org/abs/1812.06990}


\bibitem{Nakamura}
M. Nakamura, S. Masuda, and S. Nishimoto,
{\em Characterization of topological insulators based on the electronic polarization with spiral boundary conditions}\/,
Phys. Rev. B {\bf 104}, L121114 (2021).
\\\url{https://arxiv.org/abs/2109.02242}

\bibitem{AvronSeiler}
J.E. Avron and R. Seiler,
{\em Quantization of the Hall conductance for general, multiparticle Schr\"odinger Hamiltonians}\/,
Phys. Rev. Lett. {\bf 54}, 259--262 (1985).

\bibitem{Bachmann1}
S. Bachmann, A. Bols, W. De Roeck, and M. Fraas,
{\em A many-body index for quantum charge transport}\/, 
Comm. Math. Phys. {\bf 375}, 1249--1272 (2019).
\\\url{https://arxiv.org/abs/1810.07351}

\bibitem{Bachmann2}
S. Bachmann, A. Bols, W. De Roeck, and M. Fraas,
{\em Rational indices for quantum ground state sectors}\/,
J. Math. Phys. {\bf 62}, 011901 (2021).
\\\url{https://arxiv.org/abs/2001.06458}

\bibitem{BSB}
C. Bourne and H.Schulz-Baldes,
{\em On $\bbZ_2$-indices for ground states of fermionic chains}\/,
Rev. Math. Phys. {\bf 32}, 2050028 (2020).
\\\url{https://arxiv.org/abs/1905.11556v2}


\bibitem{Matsui2020}
T. Matsui,
{\em Split Property and Fermionic String Order}\/,
(preprint, 2020).
\\\url{https://arxiv.org/abs/2003.13778v2}


\bibitem{BourneOgata}
C. Bourne, and Y. Ogata, {\em The classification of symmetry protected topological phases of one-dimensional fermion systems}\/,  Forum of Mathematics, Sigma, {\bf 9}, E25 (2021). 
\\\url{doi:10.1017/fms.2021.19}

\bibitem{Ogata4}
Y. Ogata,
{\em An invariant of symmetry protected topological phases with on-site finite group symmetry for two-dimensional Fermion systems}\, 
Comm. Math. Phys. {\bf 395}, 405--457 (2022).
\\\url{https://arxiv.org/abs/2110.04672}


%bosonic

\bibitem{BachmannNachtergaele2014A}
S. Bachmann and B. Nachtergaele,
{\em On gapped phases with a continuous symmetry and boundary operators}\/,
J. Stat. Phys. {\bf 154}, 91--112 (2014).
\\\url{https://arxiv.org/abs/1307.0716}

\bibitem{Tasaki2018}
H. Tasaki,
{\em Topological phase transition and $ \mathbb{Z}_2$ index for $S=1$ quantum spin chains}\/,
Phys. Rev. Lett. {\bf 121}, 140604 (2018).
\\\url{https://arxiv.org/abs/1804.04337}

\bibitem{Ogata1}
Y. Ogata,
{\em A $ \mathbb{Z}_2$-index of symmetry protected topological phases with time reversal symmetry for quantum spin chains}\/,
preprint (2018).
Commun. Math. Phys. {\bf 374}, 705--734 (2020)
\\\url{https://arxiv.org/abs/1810.01045}"


\bibitem{Ogata2}
Y. Ogata,
{\em A $ \mathbb{Z}_2$-index of symmetry protected topological phases with reflection symmetry for quantum spin chains}\/,
Commun. Math. Phys. {\bf 385}, 1245--1272 (2021)
\\\url{https://arxiv.org/pdf/1904.01669.pdf}

\bibitem{OTT}
Y. Ogata, Y. Tachikawa, and H. Tasaki,
{\em General Lieb-Schultz-Mattis type theorems for quantum spin chains}\/,
Comm. Math. Phys.  {\bf 385}, 79--99 (2021).
\\\url{https://arxiv.org/abs/2004.06458}

\bibitem{Ogata3}
Y. Ogata,
{\em A $H^3(G,\mathbb{T})$-valued index of symmetry protected topological phases with on-site finite group symmetry for two-dimensional quantum spin systems}\/,
Forum of Mathematics, Pi (2021), Vol. 9:e13 1--62.
\\\url{https://arxiv.org/abs/2101.00426}

\bibitem{Sopenko}
N. Sopenko,
{\em An index for two-dimensional SPT states}\/,
J. Math. Phys. {\bf 62}, 111901 (2021).
\\\url{https://arxiv.org/abs/2101.00801}







\bibitem{SSH1}
W.P. Su, J.R. Schrieffer, and A.J. Heeger, {\em Solitons in Polyacetylene}\/, Phys. Rev. Lett. {\bf 42}, 1698 (1979).

\bibitem{SSH2}
W.P. Su, J.R. Schrieffer, and A.J. Heeger, {\em 
Soliton excitations in polyacetylene}\/, Phys. Rev. B {\bf 22}, 2099 (1983).

\bibitem{AOP}
J.K. Asb\'{o}th, L. Oroszl\'{a}ny, and A. P\'{a}lyi,
{\em A Short Course on Topological Insulators: Band-structure topology and edge states in one and two dimensions}\/, Lecture Notes in Physics (Springer, 2016).
\\\url{https://arxiv.org/abs/1509.02295}




\bibitem{Zak}
J. Zak, {\em Berry's phase for energy bands in solids}\/, Phys. Rev. Lett. {\bf 62}, 2747--2750 (1989).

\bibitem{Resta94}
R. Resta, Macroscopic Polarization in Crystalline Dielectrics: The Geometric Phase Approach, Rev. Mod. Phys. 66,
899 (1994).


\bibitem{Resta98}
R. Resta,
{\em The Quantum-Mechanical Position Operator in Extended Systems}\/,
Phys. Rev. Lett. {\bf 80}, 1800 (1998).
\\\url{https://arxiv.org/abs/cond-mat/9709306v1}

\bibitem{WatanabeOshikawa}
H. Watanabe and M. Oshikawa,
{\em Inequivalent Berry Phases for the Bulk Polarization}\/,
Phys. Rev. X {\bf 8}, 021065 (2018).
\\\url{https://journals.aps.org/prx/abstract/10.1103/PhysRevX.8.021065}




\bibitem{Bohm}
D. Bohm, {\em Note on a theorem of Bloch concerning possible causes of superconductivity}\/, Phys. Rev. {\bf 75},
502 (1949).

\bibitem{Watanabe}
H. Watanabe,
{\em A proof of the Bloch theorem for lattice models}\/,
J. Stat. Phys. {\bf 177}, 717--726 (2019).
\\\url{https://link.springer.com/article/10.1007%2Fs10955-019-02386-1}

\bibitem{LiebSchultzMattis1961}
E.H. Lieb, T. Schultz, and D. Mattis,
{\em Two soluble models of an antiferromagnetic chain}\/,
Ann. Phys. {\bf 16}, 407--466 (1961).

\bibitem{YOA}
M. Yamanaka, M. Oshikawa, and I. Affleck,
{\em Nonperturbative approach to Luttinger's theorem in one dimension}\/,
Phys. Rev. Lett. {\bf 79}, 1110 (1997).
\\{\tt  arXiv:cond-mat/9701141}









\bibitem{AffleckLieb1986}
I. Affleck and E.H. Lieb,
{\em A proof of part of Haldane's conjecture on spin chains}\/,
Lett. Math. Phys. {\bf 12}, 57--69 (1986).

\bibitem{Tasaki2017A}
H. Tasaki,
{\em Lieb-Schultz-Mattis theorem with a local twist for general one-dimensional quantum systems}\/,
J. Stat. Phys. {\bf 170}, 653--671 (2018).
\\\url{https://arxiv.org/abs/1708.05186}

\bibitem{NakamuraTodo2002}
M. Nakamura and S. Todo,
{\em Order Parameter to Characterize Valence-Bond-Solid States in Quantum Spin Chains}\/,
Phys. Rev. Lett. {\bf 89}, 077204 (2002).
\\\url{https://arxiv.org/abs/cond-mat/0112377}

  
\bibitem{TasakiBook}
H. Tasaki, {\em Physics and Mathematics of Quantum Many-Body Systems}\/,
Graduate Texts in Physics (Springer, 2020).


\bibitem{TasakiLSM}
H. Tasaki, {\em The Lieb-Schultz-Mattis Theorem: A Topological Point of View}\/,
in Rupert L. Frank, Ari Laptev, Mathieu Lewin, and Robert Seiringer eds. ``The Physics and Mathematics of Elliott Lieb'' vol.~2, pp.~405--446 (European Mathematical Society Press, 2022).
\\\url{https://arxiv.org/abs/2202.06243}



\bibitem{Nachtergaele2022}
B. Nachtergaele, {\em From Lieb-Robinson bounds to automorphic equivalence}\/,
in Rupert L. Frank, Ari Laptev, Mathieu Lewin, and Robert Seiringer eds. ``The Physics and Mathematics of Elliott Lieb'' vol.~2, pp.~79--92 (European Mathematical Society Press, 2022).
\\\url{https://arxiv.org/abs/2205.10460}

\bibitem{NS1}
B. Nachtergaele and R. Sims,
{\em Lieb-Robinson Bounds and the Exponential Clustering Theorem}\/,
Comm. Math. Phys. {\bf 265}, 119--130 (2006).
\\\url{https://arxiv.org/abs/math-ph/0506030}

\bibitem{HastingsKoma}
M.B. Hastings and T. Koma,
{\em Spectral Gap and Exponential Decay of Correlations}\/,
Comm. Math. Phys. {\bf 265}, 781--804 (2006).
\\\url{https://arxiv.org/abs/math-ph/0507008}

\bibitem{Zirnbauer}
M.R. Zirnbauer,
{\em Particle-Hole Symmetries in Condensed Matter}\/,
J. Math. Phys. {\bf 62},  021101 (2021).
\\\url{https://arxiv.org/abs/2004.07107v1}

\bibitem{Sompetetal}
P. Sompet, S. Hirthe, D. Bourgund, T. Chalopin, J. Bibo, J. Koepsell, P. Bojovi\'{c}, R. Verresen, F. Pollmann, G. Salomon, C. Gross, .T A. Hilker, and I. Bloch,
{\em Realising the Symmetry-Protected Haldane Phase in Fermi-Hubbard Ladders}\/,
Nature {\bf 606}, 484--488 (2022).
\\\url{https://www.nature.com/articles/s41586-022-04688-z}

\bibitem{Fuji}
Y. Fuji, F. Pollmann, and M. Oshikawa,
{\em Distinct trivial phases protected by a point-group symmetry in quantum spin chains}\/,
Phys. Rev. Lett. {\bf 114}, 177204 (2015).
\\\url{https://arxiv.org/abs/1409.8616}

\bibitem{Thouless}
D. J. Thouless,
{\em Quantization of particle transport}\/,
Phys. Rev. B {\bf 27}, 6083 (1983).

\bibitem{TasakiNext}
H. Tasaki, {\em Topological Indices, Symmetry Protected Topological Phases, Gapless Edge Excitations, Spin Pumping, and Homotopy in Quantum Spin Chains}\/, in preparation.



\end{thebibliography}
\end{document}